\documentclass[aps,prb,superscriptaddress,twocolumn,showpacs]{revtex4-2}
\usepackage{amssymb}
\usepackage{epsfig}
\DeclareMathAlphabet{\pazocal}{OMS}{zplm}{m}{n}            
\usepackage{graphicx}
\usepackage[usenames, dvipsnames]{xcolor}
\usepackage{bm}
\usepackage[normalem]{ulem}     
\usepackage{soul}
\usepackage{amsmath}
\usepackage{braket} 
\usepackage{rotating}
\usepackage{multirow} 
\usepackage{mhchem}
\DeclareMathAlphabet{\pazocal}{OMS}{zplm}{m}{n}            
\usepackage{graphicx}
\usepackage{hyperref}
\usepackage[normalem]{ulem}

\newcommand\redsout{\bgroup\markoverwith{\textcolor{red}{\rule[0.5ex]{2pt}{0.4pt}}}\ULon}
\newcommand\blacksout{\bgroup\markoverwith{\textcolor{black}{\rule[0.5ex]{2pt}{0.4pt}}}\ULon}

\newcommand{\SPhide}[1]{{}}

\begin{document}
\title{{\color{black}
Rational Control of Magnonic and Electronic Band Splittings} }        

\author{Subhadeep Bandyopadhyay}  \affiliation{Theoretical Materials Physics, Q-MAT, Université de Liège, B-4000 Sart-Tilman, Belgium}
\affiliation{Consiglio Nazionale delle Ricerche (CNR-SPIN),  Unità di Ricerca presso Terzo di Chieti, c/o Università G. D'Annunzio, I-66100 Chieti, Italy}

\author{Anoop Raj}
\affiliation{Department of Physics, Indian Institute of Technology Bombay, Mumbai 400076, India}

\author{Philippe Ghosez}
\affiliation{Theoretical Materials Physics, Q-MAT, Université de Liège, B-4000 Sart-Tilman, Belgium}

\author{Sumiran Pujari} 
\affiliation{Department of Physics, Indian Institute of Technology Bombay, Mumbai 400076, India}

\author{Sayantika Bhowal}
\email{sbhowal@iitb.ac.in}
\affiliation{Department of Physics, Indian Institute of Technology Bombay, Mumbai 400076, India}

\date{\today}

\begin{abstract}
{\color{black}
We provide a theoretical demonstration of controllable non-relativistic spin splitting in both electronic and magnonic bands via targeted structural distortions tied to specific phonon modes. Using MnF$_2$ as a model system, we identify a $d$-wave magnon band splitting between magnon modes of specific handedness, directly correlated with the non-relativistic spin splitting observed in the electronic structure. Crucially, we show that structural distortions associated with the A$_{2u}$ and A$_{1g}$ phonon modes (8.52 and 9.74 THz) modulate these splittings without altering the antiferromagnetic order. The effect originates from changes in the nonmagnetic ligand environment, highlighting the key role of lattice degrees of freedom in governing spin dynamics. Our findings establish a novel route for structure-mediated control of spin splitting, opening possibilities for tunable magnonic and spintronic functionalities in antiferromagnetic materials.
}

\end{abstract}

\maketitle

{\color{black} 
Splitting between spin-polarized bands of antiferromagnets has recently been at the center of attention. Such spin-split antiferromagnets (SSAFMs), also known as ``altermagnets", are distinguished by significant spin splitting in momentum space, even without spin-orbit coupling (SOC) or net magnetization \cite{Hayami2019,Yuan2020,Yuan2021,Smejkal2022PRX,YuanZunger2023, Guo2023, Zeng2024,Lee2024PRL,Krempask2024, Reimers2024, Aoyama2024, Lin2024, Kyo-Hoon2019, Libor2020,Libor2022Review,Paul2024,Bai2024}. By combining traits of ferromagnets and conventional antiferromagnets (AFMs), SSAFMs offer new possibilities for achieving otherwise difficult properties, such as efficient spin-current generation \cite{Naka2019, Hernandez2021, Shao2021, Bose2022,Hu2024}, spin-splitting torque \cite{Bai2022, Karube2021}, giant magnetoresistance \cite{Libor2022}, spontaneous Hall effect \cite{Libor2020, Libor2022NatRev, Helena2020, Feng2020, Betancourt2021, SmejkalAHE2022,Cheong2024, Sato2024} without any magnetization as well as unconventional superconducting properties \cite{Mazin2022,Zhu2023,Banerjee2024,Chakraborty2024,Zhang2024,Lee2024}, and enhanced thermal transport \cite{Zhou2024,Yershov2024}.

Recent theoretical predictions \cite{Libor2023,McClarty2024}, now confirmed by inelastic neutron scattering experiments \cite{Liu2024}, suggest that magnon bands in SSAFMs exhibit similar splitting to electronic bands, resulting in non-degenerate magnon modes with specific handedness, referred to as ``chiral'' magnons \cite{Libor2023}. The handedness can be thought of in analogy with the polarization of photons as quanta of electromagnetic vector fields, since magnons are quanta of waves made out of electron spins, which are also vectorial in nature \footnote{We will stick to handedness in the rest of the article}. 
This development brings us closer to realizing advanced magnonic devices that rely on magnon-generated spin currents in AFMs, functioning without the need for external magnetic fields to generate the splitting \cite{Cui2023}.

In this work, we show the control of splitting in spin-polarized electronic and magnonic bands by controlling the structural distortion associated with specific phonon modes with frequencies in the terahertz (THz) regime without changing the AFM order. We illustrate our findings using MnF$_2$ as a representative material for SSAFMs. Our interest in MnF$_2$ stems from recent discoveries in this nearly century-old material \cite{deHaas1940,Seehra1984,Yamani2010}, which highlight unusual spin splitting \cite{Yuan2020}, the presence of ferroic ordering of higher-order magnetic octupoles \cite{BhowalSpaldin2024,Costa1989}, and ongoing efforts \cite{Mashkovich2021, Disa2020, Metzger2024, Dubrovin2024} to manipulate its magnetism via structural modifications. 

Our key findings are threefold. First, we demonstrate that the magnon bands in MnF$_2$ exhibit a $d$-wave splitting, similar to its electronic bands. We identify the wave vector where not only is the splitting maximum, but also the magnon bands possess a pure left and right-handedness, which can be crucial for detection using polarized neutron or X-ray scattering techniques. Second, we identify \textcolor{black}{structural distortions that are related to} two stable $\Gamma$-point phonon modes in MnF$_2$ and show that by jointly controlling these \textcolor{black}{distortions}, it is possible to manipulate the energy splitting between the magnon bands. In practice, such structural control can be achieved, for example, using an external electric field, optical pulses, or strain engineering. Third, we reveal a striking correlation between electronic and magnonic bands, enabling the control of spin splitting in the electronic bands, leading to a transition from SSAFM to a conventional AFM with no spin splitting.

Our work advances the current understanding of the interplay between structural geometry and magnetism in SSAFMs. We go beyond symmetry considerations to show that the local geometry of nonmagnetic ligand atoms plays a crucial role in inducing anisotropic magnetization density in the ground state, leading to direction-dependent exchange coupling between magnetic atoms. While the anisotropic magnetization density, characterized by ferroic magnetic octupoles \cite{BhowalSpaldin2024}, is essential for spin splitting in electronic bands, we analytically demonstrate that magnon band splitting arises from the direction-dependent exchange coupling. Thus, by tuning the positions of nonmagnetic ligand atoms following specific phonon modes, we can simultaneously control the electronic and magnonic spin splittings.

Our work takes a new step in the direction of the recent efforts \cite{Duan2025, Gu2025,Libor2025} to control non-relativistic spin splitting (NRSS) by introducing a universal approach that extends beyond (anti-) ferroelectric materials. We show that NRSS can be tuned by manipulating the ligand ion environment through specific structural phonon mode distortions. This phonon-assisted control also opens a pathway for ultrafast manipulation of next-generation magnonic devices \cite{Kirilyuk2010}.

}

\begin{figure}[t]
\centering
\includegraphics[width=\columnwidth]{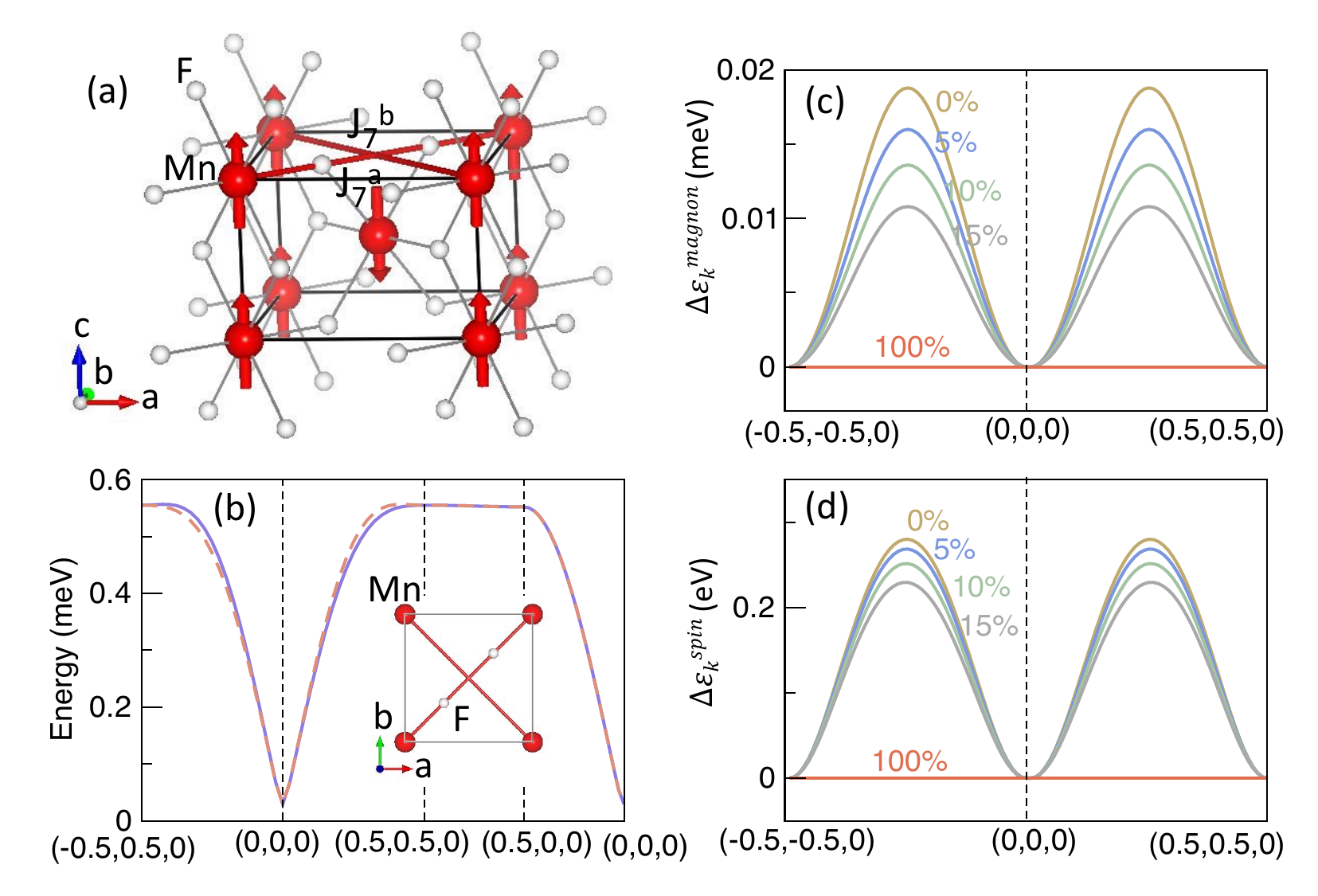}
 \caption{ Control of magnon and electronic spin splitting in MnF$_2$ by tuning 
 A$_{2u}$ and A$_{1g}$ \textcolor{black}{distortions} jointly. (a) Crystal structure of MnF$_2$, indicating the Mn ions on the $ab$ plane corresponding to the $J_7$ exchange coupling. {\color{black}The bonds indicate the Mn-Mn interaction paths along $[110]$ ($J_7^a$) and $[1\bar{1}0]$ ($J_7^b$) directions with and without the F ions sitting in between, respectively. The arrows indicate the direction of the Mn spin moments in the ground state of MnF$_2$.} (b) Computed magnon spectrum of MnF$_2$. The $\gamma_k$ term, identical for both branches, has been scaled by a factor of 0.1 to make the splitting more apparent. {\color{black} The inset shows the  $z = 0$ plane of MnF$_2$, highlighting the presence and absence of F-ions along the $J_7^a$ and $J_7^b$ exchange paths.} (c) The variation in magnon splitting as a function of 
 combined distortion of phonon modes, A$_{2u}$ and A$_{1g}$, along specific directions in the momentum space. (d) The corresponding NRSS in the electronic band structure for the second-most top pairs of valence bands along the same $k$ path for different amplitudes of the distortions.
 }
 \label{fig1}
 \end{figure}

{\it Magnon spectra in MnF$_2$-}
We begin by demonstrating the magnon band splitting in MnF$_2$. MnF$_2$ crystallizes in the rutile structure ($P4_2/mnm$) (see Fig. \ref{fig1}a), with spin moments aligned along the [001] direction and antiparallel between Mn ions \cite{deHaas1940,Seehra1984,Yamani2010}. This magnetic ordering breaks global time-reversal symmetry ($\cal T$), resulting in spin splitting in the ground state electronic band structure \cite{Yuan2020, BhowalSpaldin2024}. To investigate whether the magnon bands exhibit a similar splitting, we first construct the relevant spin model by calculating the Heisenberg exchange interactions between Mn$^{2+}$ ions ($S=5/2$). These interactions are computed by mapping the total energy obtained from DFT+$U$ \cite{Bloch1994,Kresse1999, Kresse1993,Kresse1996} calculations to a Heisenberg spin model for four different collinear magnetic configurations \cite{Xiang2011, Sabani2020, SM}. 

{\color{black} Our computed exchange interactions ($J_i$) up to the seventh nearest neighbors (NNs) ($i=1,7$) [See Figs. 1 and 2 of the Supplemental Materials (SM) \cite{SM}] show that  
the magnitude of the intra-sublattice coupling $J_7$, corresponding to the seventh NN Mn-Mn distance ($d_7 = 7.03$ \AA), varies with direction ($[110]$ vs. $[1\bar{1}0]$)}. This direction-dependent exchange interaction arises from the crystal geometry. {\color{black} As illustrated in Fig. \ref{fig1}a and the inset of Fig. \ref{fig1}b, two F atoms at $(x,x,0)$ and $(-x,-x,0)$ [or $(-x+\tfrac{1}{2},x+\tfrac{1}{2},\tfrac{1}{2})$ and $(x+\tfrac{1}{2},-x+\tfrac{1}{2},\tfrac{1}{2})$] lie along the $[110]$ ($[1\bar{1}0]$) direction, but not along $[1\bar{1}0]$ ($[110]$), of the $J_7$ exchange path connecting the corner (central) Mn ions.} This leads to stronger coupling along the former and weaker coupling along the latter. We refer to these exchange couplings as $J_7^a$ and $J_7^b$, respectively. A similar direction-dependent exchange has been reported for the iso-structural compound RuO$_2$ \cite{Libor2023}. {\color{black} However, the further-neighbor coupling  $J_7$ in MnF$_2$ is much weaker than in RuO$_2$, likely due to the itinerant magnetism considered in the latter \cite{Libor2023, note,Smolyanyuk2024, Hiraishi2024,Philipp2024}.
}

{\color{black} We further compute the uniaxial single-ion anisotropy $D_c$ from the energy difference between AFM configurations with spins along $\hat x$ and $\hat z$ in the presence of SOC.  Using the computed exchange couplings $J_{ij}$ (up to 7th NN) and $D_c$, we construct the spin Hamiltonian for MnF$_2$,  ${\cal H}_{\rm spin} = \sum_{ij} J_{ij} \vec{S}_i \cdot \vec{S}_j + \sum_{i} D_c (S_i^z)^2$}. 
Through analytical spin-wave calculations, using the Holstein-Primakoff transformation \cite{HolsteinPrimakoff1940} and the Fourier transformation, we obtain the Hamiltonian in terms of the bosonic operators $a_k$ and $b_k$ in the reciprocal space as 
\begin{equation}
    H_{k} = \sum_{k} [A_{k} a_{k}^{\dagger}a_{k} + B_{k}b_{k}^{\dagger}b_{k}+C_{k}a_{k}b_{-k}+C_{k}^{*} a_{k}^{\dagger}b_{-k}^{\dagger}]
\end{equation} 
{\color{black} Here, $a_k$ ($a_k^{\dagger}$) and $b_k$ ($b_k^{\dagger}$) are the Fourier transforms of the bosonic operators $a_i$ ($a_i^{\dagger}$) and $b_i$ ($b_i^{\dagger})$, respectively. The functions $A_k, B_k,$ and $C_k$ are periodic in momentum $\vec k$ and explicitly depend on the exchange parameters $J_i$ (see SM \cite{SM} for details).} We then diagonalize the Hamiltonian using Bogoliubov transformation. The resulting Hamiltonian in terms of the
magnon operators of $\alpha$ and $\beta$ modes reads as:
\begin{equation}
    H_{k} = \sum_{k} [\epsilon_{\alpha}(k) \alpha_{k}^{\dagger}\alpha_{k} + \epsilon_{\beta}(k) \beta_{k}^{\dagger}\beta_{k}].
\end{equation}
Here, $\epsilon_{\alpha, \beta}(k) = \frac{1}{2}(\pm\Delta \varepsilon_k + \gamma_k)$, with {\color{black} $\gamma_k = [(A_{k}+B_{k} - 4D_c)^{2} - 4C_{k}^{2}]^{1/2}$} and the energy splitting $\Delta \varepsilon_k^{magnon}$ between the two magnon modes is
\begin{equation}\label{splitting}
    \Delta \varepsilon_k^{magnon} = \epsilon_{\alpha}(k)- \epsilon_{\beta}(k) = 4 (J_{7}^{b}-J_{7}^{a}) \sin (k_xa) \sin (k_ya).
\end{equation} 
This constitutes an analytical demonstration of the splitting between the magnon bands (see Fig. \ref{fig1}b) occurring due to the direction-dependent exchange coupling $J_7$. It is also clear from Eq. (\ref{splitting}) that $\Delta \varepsilon_k^{magnon}$ switches sign under $C_{4z}$ rotation of the wave vector $\vec k$, leading to a $d$-wave splitting similar to the NRSS in the electronic bands. {\color{black} 
We note that the calculated magnon band splitting, \( \Delta \varepsilon_k^{\text{magnon}} \), is small in magnitude ($ \sim 0.02 \, \text{meV}$) for MnF\(_2\). A recent \emph{ unpolarized} inelastic neutron scattering experiment, with a resolution of approximately 0.1 meV \cite{Morano2024}, did not detect this splitting and concluded that it is absent, suggesting the need for techniques such as polarized neutrons or circularly polarized photons, which are sensitive to the handedness of magnons.  
Interestingly, our calculations show that at \( \vec{k} \equiv (\pm \frac{\pi}{2a}, \pm\frac{\pi}{2a}, \pm\frac{\pi}{c}) \), the splitting \( \Delta \varepsilon_k^{\text{magnon}} \) is not only maximum but also the eigenvectors take the form \( [1~~ 0]^{\rm T} \) and \( [0~~ 1]^{\rm T} \), exhibiting {\it pure} left- and right-handedness. This indicates a stronger coupling of magnon bands to circularly polarized light or polarized neutrons at these momentum points.  

}

\begin{figure}[t]
\centering
\includegraphics[width=\columnwidth]{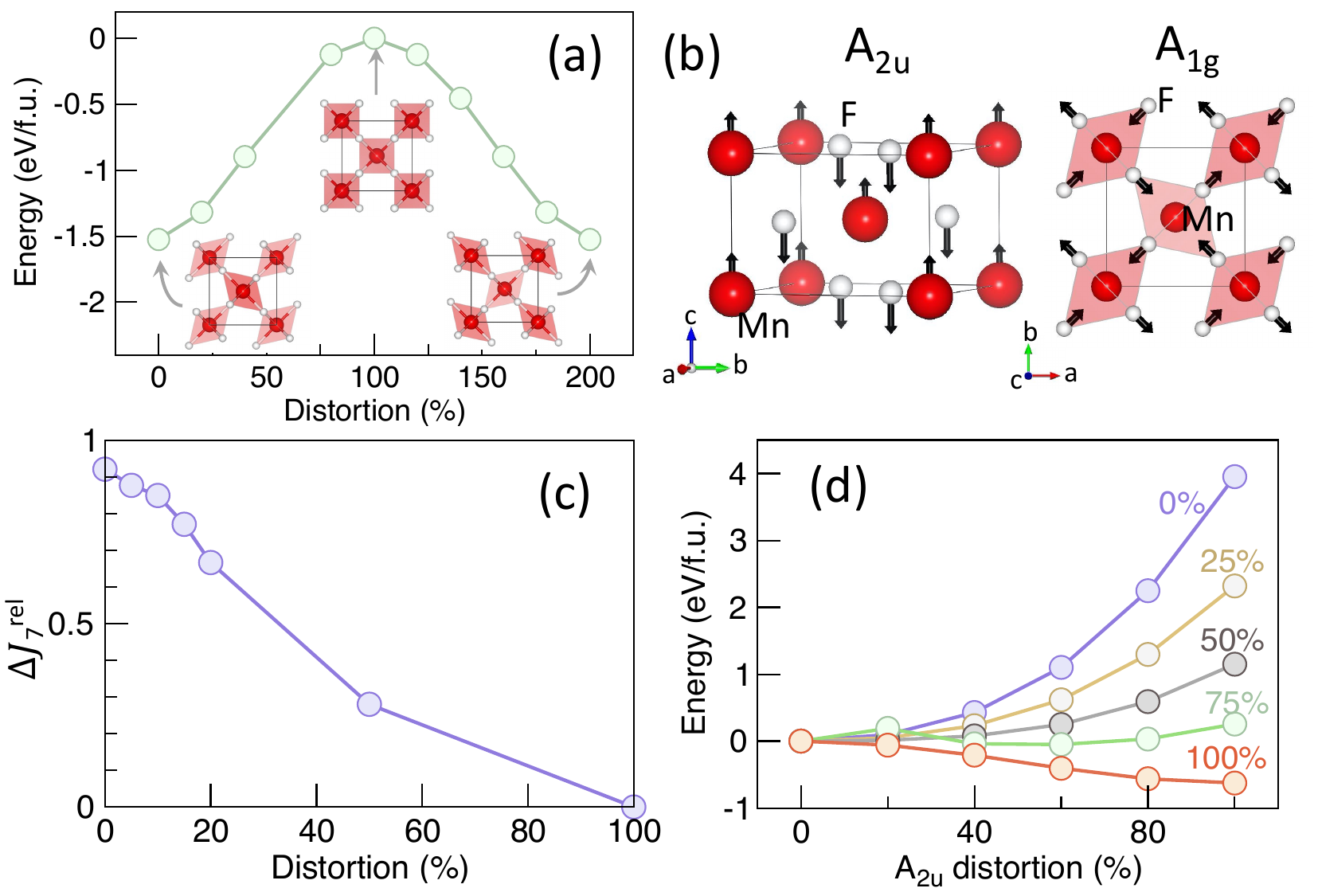}
 \caption{ The effect of structural distortion. (a) The variation in the total energy as a function of distortion amplitude. The $0 \%, 100 \%$, and $200 \%$ distortions correspond to the actual crystal structure, the higher energy structure with zero spin splitting, and the opposite structural domain with opposite spin splitting. (b)  Pictorial depiction of A$_{2u}$ ({\it left}) and A$_{1g}$ ({\it right}) phonon modes. The black arrows indicate the displacements of the Mn and F ions in the unit cell. (c) The dependence of $\Delta J_7^{\rm rel}$ on the amplitude of the combined control of the phonon modes A$_{2u}$ and A$_{1g}$. (d) The total energy as a function of A$_{2u}$ distortion for different amplitudes of A$_{1g}$ distortions, showing that the structure with stronger A$_{2u}$ distortion also energetically favors stronger A$_{1g}$ distortion. 
 }
 \label{fig2}
 \end{figure}

{\it Control of magnon splitting-}
Since the magnon splitting $\Delta \varepsilon_k^{magnon}$ originates from the inequivalent $J_7$ exchange couplings, $\Delta \varepsilon_k^{magnon}$ can be tuned by controlling the difference $\Delta J_7= (J_7^b-J_7^a)$. Furthermore, since $\Delta J_7$ results from the inequivalent F environment around the Mn ions for the different exchange paths corresponding to $J_7^a$ and $J_7^b$,  we can control $\Delta J_7$ and hence the magnon splitting by manipulating the position of the F atoms. 

With this idea, we construct a higher-energy structure with zero spin splitting, corresponding to the space group $P4_2/nnm$, by linearly interpolating the two structural domains of MnF$_2$ (see the inset of Fig. \ref{fig2}a) with opposite spin splitting \cite{BhowalSpaldin2024}. In this structure, the F atoms are shifted from the plane containing the Mn atoms such that no F atoms are present between the seventh neighbor Mn atoms. 
We emphasize that our choice of the higher-energy structure with zero spin splitting is not unique. However, it is guided by the fact that the distortion linked to  this reference $P4_2/nnm$ structure 
can be represented in terms of the two stable $\Gamma$ optical phonon modes corresponding to the $A_{2u}$ and $A_{1g}$ irreducible representations (IRs) of the ground state $P4_2/mnm$ structure. 
We identify these phonon modes by explicitly carrying out the phonon calculations for the ground state $P4_2/mnm$ structure using the finite difference method as implemented within the Phonopy ~\cite{Phonopy} and then projecting the distorted $P4_2/nnm$ structure on the $\Gamma$ point phonon eigenvectors of the $P4_2/mnm$ structure \cite{Isodistort, Isodistort1}.

We note that the polar distortion $A_{2u}$ is infrared active \cite{Dubrovin2024}, describing the displacements of the Mn and F atoms in the opposite direction along the $c$ axis (see Fig. \ref{fig2}b) such that it lowers the symmetry of the structure to $P4_2nm$. 
In contrast, the A$_{1g}$ phonon mode is isosymmetric and Raman-active. The A$_{1g}$ phonon mode describes the displacement of the F atoms on the $ab$ plane, as depicted in Fig. \ref{fig2}b, which resembles the $Q_{2z}$ Jahn-Teller distortion \cite{SB_TMO} of the MnF$_6$ octahedra. 
The computed frequencies of the A$_{2u}$ and A$_{1g}$ phonon modes are, respectively, 284 cm$^{-1}$ (8.52 THz) and 325 cm$^{-1}$ (9.74 THz). 

We now show that magnon splitting indeed can be tuned by jointly controlling the two phonon mode distortions. For this, we construct several intermediate structures between the ground state structure $P4_2/mnm$ and the higher energy structure $P4_2/nnm$ by varying the amplitudes of the atomic distortions as $\beta(Q_{{\rm A}_{2u}}+Q_{{\rm A}_{1g}})$. Here, $Q_{{\rm A}_{2u}}$ and $Q_{{\rm A}_{1g}}$ are the atomic distortions corresponding to the A$_{2u}$ and A$_{1g}$ phonon modes respectively \footnote{ $Q_{A_{2u}}$ and $Q_{A_{1g}}$ are normalized to their amplitudes at $P4_2/nnm$ structure}, and $\beta$ varies between 0 to 1 which correspond to 0\% ($P4_2/mnm$) and 100\% ($P4_2/nnm$) of atomic distortions respectively. 
We then compute the variation of the exchange interactions across these intermediate structures. The resulting change $\Delta J_7^{\rm rel}$ in the seventh NN coupling,  $\Delta J_7^{\rm rel}=|\frac{\Delta J_7}{(J_7^a+J_7^b)}|$, is shown in Fig. \ref{fig2}c. As we see from the plot, $\Delta J_7^{\rm rel}$ gradually decreases with an increase in the distortion and eventually vanishes at the higher energy $P4_2/nnm$ structure (See Fig. \ref{fig2}c){\color{black}, consistent with the equivalent $J_7$ exchange paths.} Consequently, we can tune the magnon splitting $\Delta \varepsilon_k^{magnon}$ by controlling the amplitude of the structural distortion, as depicted in Fig. \ref{fig1}c. We note that with increasing distortion, $J_2$ and $J_5$ exchange couplings also gradually become inequivalent \cite{SM}. However, both being inter-sublattice coupling, {\color{black} the corresponding changes only affect the $\gamma_k$ term in the magnon band dispersion and} do not contribute to the splitting of the magnon bands {\color{black} $\Delta \varepsilon_k^{magnon}$ \cite{SM}}. 

{\color{black} We emphasize that the magnon splitting control discussed here arises from changes in the nonmagnetic environment around the Mn ions, without changing the AFM order, as confirmed by our computed exchange couplings \cite{SM}. This ``magnon splitting" is fundamentally different from the splitting reported in earlier works \cite{Hioki2022, Liu2021, Mai2021, Cui2023,Metzger2024}, which arises from magnon-phonon hybridization and appears as an anti-crossing gap between magnon and phonon bands. In contrast, our work explores how structural distortions—driven by specific phonon modes—affect electronic and magnonic bands within the Born-Oppenheimer approximation, where these degrees of freedom respond instantaneously to changes in ionic positions.} 

\textcolor{black} {To better rationalize the joint control of the A$_{2u}$ and A$_{1g}$  distortions, we
perform a low-order Taylor expansion of the Born-Oppenheimer energy $U$ around the ground-state $P4_2/mnm$ structure in terms of their respective amplitudes $Q_{{\rm A}_{2u}}$ and $Q_{{\rm A}_{1g}}$ \cite{symm,Invariants, Invariants2}:}
\begin{eqnarray}
    U &\approx & \alpha_1Q_{{\rm A}_{1g}}^2+  \alpha_2Q_{{\rm A}_{1g}}^4+ \beta_1Q_{{\rm A}_{2u}}^2+  \beta_2Q_{{\rm A}_{2u}}^4
     \nonumber\\
   & + & \lambda_1 Q_{{\rm A}_{2u}}^2Q_{{\rm A}_{1g}} + \lambda_2 Q_{{\rm A}_{2u}}^2Q_{{\rm A}_{1g}}^2+ O(Q^5).
   \label{freeenergy}
\end{eqnarray}
{\color{black}  By fitting DFT total energies of a set of structures with
different combinations of $Q_{{\rm A}_{2u}}$ and $Q_{{\rm A}_{1g}}$ with Eq. (\ref{freeenergy}), we extract the values of the coefficients:  $\alpha_1= 1.061 $,   $\alpha_2= 0.346 $, $\beta_1= 2.944$, $\beta_2= 0.907$, $\lambda_1 =-6.253$ and $\lambda_2=1.812$ eV/f.u. (see SM\cite{SM} for details).
Using Eq.~(\ref{freeenergy}), we then compute the optimal values of $Q_{{\rm A}_{1g}}$ for a given $Q_{{\rm A}_{2u}}$, which interestingly vary as 
$Q_{{\rm A}_{1g}} \approx Q_{{\rm A}_{2u}}$, supporting the approximation we made previously.

In practice, infrared-active modes can be tuned using optical pulses, as shown in CoF$_2$ \cite{Disa2020}, or via an external electric field ${\cal E}_z$. To estimate the required field strength, we have added a term $-\Omega {\cal E}_z P_z$ to Eq.~(\ref{freeenergy}), with $P_z \propto Q_{{\rm A}_{2u}}$ and $\Omega$ as the unit cell volume (see SM \cite{SM}). We find that ${\cal E}_z$ of 6.5, 11.5, and 15 MV/cm induce 5\%, 10\%, and 15\% of $Q_{{\rm A}_{2u}}$ distortions, modifying the NRSS by 4\%, 10\%, and 18\%, respectively. While these fields are large, they are potentially achievable \cite{Kyung_2023, Varignon_2016, Bellaiche_2014}. This serves as a proof of concept, and the effect could be enhanced with a softer $A_{2u}$ mode, possibly via alternative compounds \cite{Bandyopadhyay2025} or strain engineering \cite{SM}.
}

 {\it Connection to spin splitting in the electronic bands-}
To investigate the impact of joint control of A$_{2u}$ and A$_{1g}$ structural distortions on the electronic bands, we further analyze the corresponding electronic bands for intermediate distortions. The results are shown in Fig. \ref{fig1}d. As evident from the figure, the spin splitting between electronic bands decreases with increasing distortion and, eventually, vanishes for the higher-energy $P4_2/nnm$ structure, similar to what we observe with the magnon bands. 
This indicates a correlation between the splitting in the electronic and magnonic bands. 

\begin{figure}[t]
\centering
\includegraphics[width=\columnwidth]{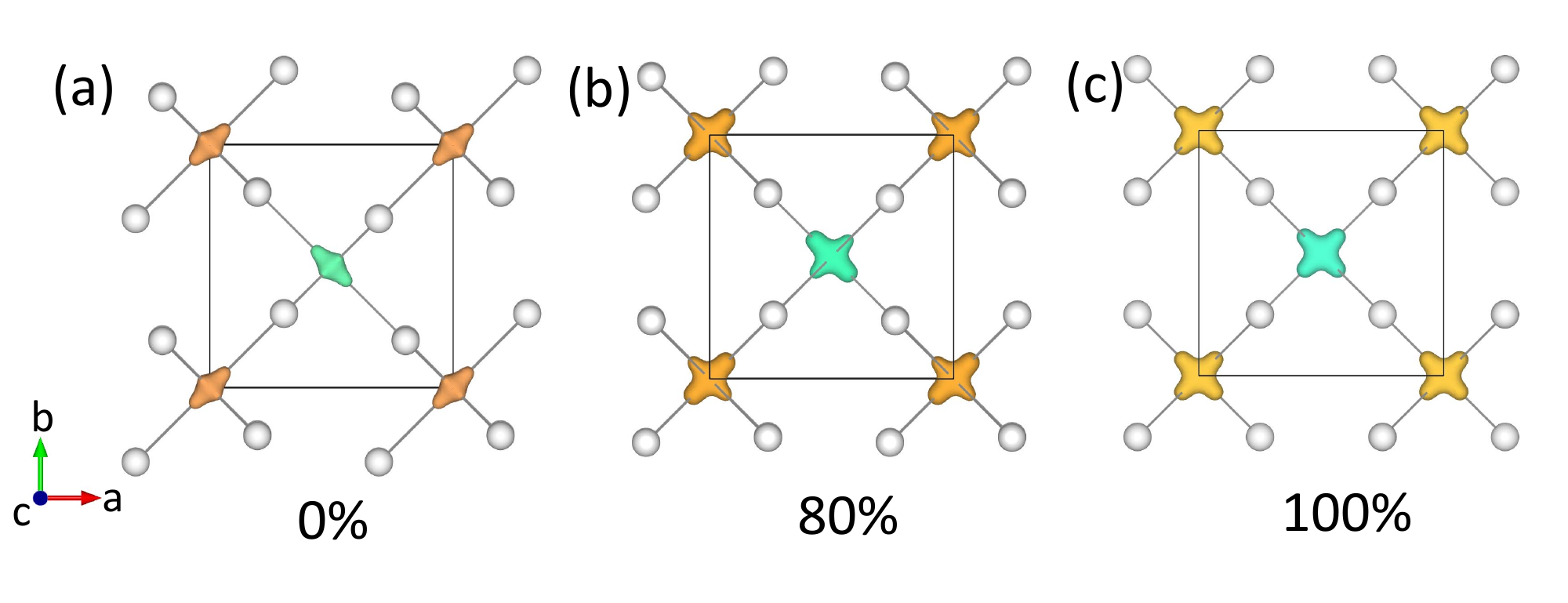}
 \caption{The computed magnetization densities for three different distortion amplitudes (a) 0\%, (b) 80\%, and (c) 100\%. The isosurface value is set to 0.1 for all the plots. Different colors of the magnetization density denote the opposite spin polarization around the corner and center Mn ions. The white balls represent the F atoms. 
 }
 \label{fig3}
 \end{figure}

To explore the origin of the connection between electronic and magnonic bands, we compute the band-decomposed magnetization density for the top pair of valence bands for structures with 0\%, 80\%, and 100\% distortions. Here, 0\% corresponds to the ground-state structure, while 100\% refers to the $P4_2/nnm$ structure with zero spin splitting. As seen in Figs. \ref{fig3} a-c, the anisotropy in the magnetization density decreases with increasing distortion. For 0\% distortion, the magnetization density around the Mn ions is highly anisotropic, taking on an elliptical shape, with the long axis aligned along the Mn-Mn bond corresponding to the seventh neighbor, with two F atoms between them. This anisotropic magnetization is linked to the direction-dependent $J_7$ exchange coupling as the larger magnetization density implies a stronger interaction along that particular Mn-Mn bond.

Interestingly, the anisotropic magnetization density, described by the magnetic octupole, has previously been shown to play a key role in the spin splitting of electronic bands \cite{BhowalSpaldin2024}. The computed magnitude of the ferroically ordered atomic site magnetic octupole component ${\cal O}_{32^-} = -2.5 \times 10^{-3} \mu_B$ per Mn atom. As distortion increases, the F atoms move away from the Mn-Mn bond, and the magnetization density shifts to form two ellipsoids along the Mn-Mn bonds associated with $J_7$, thus reducing $\Delta J_7^{\rm rel}$. At 100\% distortion, the anisotropy in the magnetization density is minimized, and our multipole calculations \cite{Cricchio2010, Granas2012, Spaldin2013, SM} show the absence of ferroic magnetic octupoles with ${\cal O}_{32^-}=0$. {\color{black} This demonstrates a transition from an ``altermagnetic" state with ferromagnetooctupole to a conventional antiferromagnetic state with no ferroic magnetic octupole and NRSS}. 
Correspondingly, $\Delta J_7^{\rm rel}$ also vanishes, resulting in an absence of splitting between the magnon bands, revealing the origin of the connection between splitting in the electronic and magnonic bands.

The structural manipulation of electronic and magnonic splitting demonstrated in our study extends beyond MnF$_2$ and related difluorides. {\color{black} Recent predictions \cite{Duan2025, Gu2025, Libor2025, Bandyopadhyay2025} of NRSS tuning via electric field or strain further reinforce our findings.} The proposed concept is general and also applies to other SSAFMs, including those exhibiting $g$-wave spin splitting with anisotropic magnetization density, characterized by magnetic triakontadipoles \cite{Verbeek2024, Hoyer2025} rather than magnetic octupoles. While the specific phonon modes may vary across different materials, it remains feasible to control the positioning of nonmagnetic ligand atoms by carefully analyzing the relevant phonon distortion modes, thus enabling the regulation of splitting in both magnon and electronic bands. 
 
To summarize, our work uncovers the intricate correlation between phonons, electrons, and magnons in SSAFMs, by explicitly demonstrating the control of the dispersion relation for the last two by tuning the former. The proposed phonon-assisted manipulation of the magnon and electronic splitting may be probed using time-resolved magneto-optical Kerr measurements and time-resolved resonant inelastic scattering measurements. {\color{black} Interestingly, a recent study on CoF$_2$ demonstrates the coupling between magnons and infrared-active polar phonons through dynamic modulation of magnetic exchange interactions using far-infrared and dielectric spectroscopy techniques \cite{Dubrovin2024} — going beyond the conventional avoided crossing between phonons and magnons. These results offer a promising experimental approach to test and validate the control mechanism we propose for MnF$_2$.}
We hope that our work will open up a new paradigm for ultra-fast manipulation of SSAFMs, motivating future research along these directions.

\section*{Acknowledgements}
We thank Abhishek Nag, Urs Staub, and Nicola A. Spaldin for stimulating discussions.
SBa acknowledges use of the CECI supercomputer facilities funded by the F.R.S-FNRS (Grant No. 2.5020.1) and of the Tier-1 supercomputer of the Fédération Wallonie-Bruxelles funded by the Walloon Region (Grant No. 1117545). AR acknowledges the financial support provided by CSIR-HRDG, India, in the form of a Senior Research Fellowship (SRF). SP acknowledges financial support from SERB, DST, Govt. of India (Grant No. MTR/2022/000386) and partially by Grant No. CRG/2021/003024. SBh thanks National Supercomputing Mission for providing computing resources
of ‘PARAM Porul’ at NIT Trichy, implemented by C-DAC and supported by the Ministry
of Electronics and Information Technology (MeitY) and Department of Science and
Technology, Government of India and acknowledges funding support from the Industrial Research and Consultancy Centre (IRCC) Seed Grant (RD/0523-IRCCSH0-018) and the INSPIRE research grant (project code RD/0124-DST0030-002).

\bibliography{SB}

\providecommand{\noopsort}[1]{}\providecommand{\singleletter}[1]{#1}%
\begin{thebibliography}{94}%
\makeatletter
\providecommand \@ifxundefined [1]{%
 \@ifx{#1\undefined}
}%
\providecommand \@ifnum [1]{%
 \ifnum #1\expandafter \@firstoftwo
 \else \expandafter \@secondoftwo
 \fi
}%
\providecommand \@ifx [1]{%
 \ifx #1\expandafter \@firstoftwo
 \else \expandafter \@secondoftwo
 \fi
}%
\providecommand \natexlab [1]{#1}%
\providecommand \enquote  [1]{``#1''}%
\providecommand \bibnamefont  [1]{#1}%
\providecommand \bibfnamefont [1]{#1}%
\providecommand \citenamefont [1]{#1}%
\providecommand \href@noop [0]{\@secondoftwo}%
\providecommand \href [0]{\begingroup \@sanitize@url \@href}%
\providecommand \@href[1]{\@@startlink{#1}\@@href}%
\providecommand \@@href[1]{\endgroup#1\@@endlink}%
\providecommand \@sanitize@url [0]{\catcode `\\12\catcode `\$12\catcode
  `\&12\catcode `\#12\catcode `\^12\catcode `\_12\catcode `\%12\relax}%
\providecommand \@@startlink[1]{}%
\providecommand \@@endlink[0]{}%
\providecommand \url  [0]{\begingroup\@sanitize@url \@url }%
\providecommand \@url [1]{\endgroup\@href {#1}{\urlprefix }}%
\providecommand \urlprefix  [0]{URL }%
\providecommand \Eprint [0]{\href }%
\providecommand \doibase [0]{https://doi.org/}%
\providecommand \selectlanguage [0]{\@gobble}%
\providecommand \bibinfo  [0]{\@secondoftwo}%
\providecommand \bibfield  [0]{\@secondoftwo}%
\providecommand \translation [1]{[#1]}%
\providecommand \BibitemOpen [0]{}%
\providecommand \bibitemStop [0]{}%
\providecommand \bibitemNoStop [0]{.\EOS\space}%
\providecommand \EOS [0]{\spacefactor3000\relax}%
\providecommand \BibitemShut  [1]{\csname bibitem#1\endcsname}%
\let\auto@bib@innerbib\@empty
\bibitem [{\citenamefont {Hayami}\ \emph {et~al.}(2019)\citenamefont {Hayami},
  \citenamefont {Yanagi},\ and\ \citenamefont {Kusunose}}]{Hayami2019}%
  \BibitemOpen
  \bibfield  {author} {\bibinfo {author} {\bibfnamefont {S.}~\bibnamefont
  {Hayami}}, \bibinfo {author} {\bibfnamefont {Y.}~\bibnamefont {Yanagi}},\
  and\ \bibinfo {author} {\bibfnamefont {H.}~\bibnamefont {Kusunose}},\
  }\bibfield  {title} {\bibinfo {title} {Momentum-dependent spin splitting by
  collinear antiferromagnetic ordering},\ }\href
  {https://doi.org/10.7566/JPSJ.88.123702} {\bibfield  {journal} {\bibinfo
  {journal} {J. Phys. Soc. Jpn.}\ }\textbf {\bibinfo {volume} {88}},\ \bibinfo
  {pages} {123702} (\bibinfo {year} {2019})}\BibitemShut {NoStop}%
\bibitem [{\citenamefont {Yuan}\ \emph {et~al.}(2020)\citenamefont {Yuan},
  \citenamefont {Wang}, \citenamefont {Luo}, \citenamefont {Rashba},\ and\
  \citenamefont {Zunger}}]{Yuan2020}%
  \BibitemOpen
  \bibfield  {author} {\bibinfo {author} {\bibfnamefont {L.-D.}\ \bibnamefont
  {Yuan}}, \bibinfo {author} {\bibfnamefont {Z.}~\bibnamefont {Wang}}, \bibinfo
  {author} {\bibfnamefont {J.-W.}\ \bibnamefont {Luo}}, \bibinfo {author}
  {\bibfnamefont {E.~I.}\ \bibnamefont {Rashba}},\ and\ \bibinfo {author}
  {\bibfnamefont {A.}~\bibnamefont {Zunger}},\ }\bibfield  {title} {\bibinfo
  {title} {Giant momentum-dependent spin splitting in centrosymmetric low-$z$
  antiferromagnets},\ }\href {https://doi.org/10.1103/PhysRevB.102.014422}
  {\bibfield  {journal} {\bibinfo  {journal} {Phys. Rev. B}\ }\textbf {\bibinfo
  {volume} {102}},\ \bibinfo {pages} {014422} (\bibinfo {year}
  {2020})}\BibitemShut {NoStop}%
\bibitem [{\citenamefont {Yuan}\ \emph {et~al.}(2021)\citenamefont {Yuan},
  \citenamefont {Wang}, \citenamefont {Luo},\ and\ \citenamefont
  {Zunger}}]{Yuan2021}%
  \BibitemOpen
  \bibfield  {author} {\bibinfo {author} {\bibfnamefont {L.-D.}\ \bibnamefont
  {Yuan}}, \bibinfo {author} {\bibfnamefont {Z.}~\bibnamefont {Wang}}, \bibinfo
  {author} {\bibfnamefont {J.-W.}\ \bibnamefont {Luo}},\ and\ \bibinfo {author}
  {\bibfnamefont {A.}~\bibnamefont {Zunger}},\ }\bibfield  {title} {\bibinfo
  {title} {Prediction of low-z collinear and noncollinear antiferromagnetic
  compounds having momentum-dependent spin splitting even without spin-orbit
  coupling},\ }\href {https://doi.org/10.1103/PhysRevMaterials.5.014409}
  {\bibfield  {journal} {\bibinfo  {journal} {Phys. Rev. Materials}\ }\textbf
  {\bibinfo {volume} {5}},\ \bibinfo {pages} {014409} (\bibinfo {year}
  {2021})}\BibitemShut {NoStop}%
\bibitem [{\citenamefont {\ifmmode~\check{S}\else \v{S}\fi{}mejkal}\ \emph
  {et~al.}(2022{\natexlab{a}})\citenamefont {\ifmmode~\check{S}\else
  \v{S}\fi{}mejkal}, \citenamefont {Sinova},\ and\ \citenamefont
  {Jungwirth}}]{Smejkal2022PRX}%
  \BibitemOpen
  \bibfield  {author} {\bibinfo {author} {\bibfnamefont {L.}~\bibnamefont
  {\ifmmode~\check{S}\else \v{S}\fi{}mejkal}}, \bibinfo {author} {\bibfnamefont
  {J.}~\bibnamefont {Sinova}},\ and\ \bibinfo {author} {\bibfnamefont
  {T.}~\bibnamefont {Jungwirth}},\ }\bibfield  {title} {\bibinfo {title}
  {Beyond conventional ferromagnetism and antiferromagnetism: A phase with
  nonrelativistic spin and crystal rotation symmetry},\ }\href
  {https://doi.org/10.1103/PhysRevX.12.031042} {\bibfield  {journal} {\bibinfo
  {journal} {Phys. Rev. X}\ }\textbf {\bibinfo {volume} {12}},\ \bibinfo
  {pages} {031042} (\bibinfo {year} {2022}{\natexlab{a}})}\BibitemShut
  {NoStop}%
\bibitem [{\citenamefont {Yuan}\ and\ \citenamefont
  {Zunger}(2023)}]{YuanZunger2023}%
  \BibitemOpen
  \bibfield  {author} {\bibinfo {author} {\bibfnamefont {L.-D.}\ \bibnamefont
  {Yuan}}\ and\ \bibinfo {author} {\bibfnamefont {A.}~\bibnamefont {Zunger}},\
  }\bibfield  {title} {\bibinfo {title} {Degeneracy removal of spin bands in
  collinear antiferromagnets with non-interconvertible spin-structure motif
  pair},\ }\href {https://doi.org/https://doi.org/10.1002/adma.202211966}
  {\bibfield  {journal} {\bibinfo  {journal} {Advanced Materials}\ }\textbf
  {\bibinfo {volume} {35}},\ \bibinfo {pages} {2211966} (\bibinfo {year}
  {2023})}\BibitemShut {NoStop}%
\bibitem [{\citenamefont {Guo}\ \emph {et~al.}(2023)\citenamefont {Guo},
  \citenamefont {Liu}, \citenamefont {Janson}, \citenamefont {Fulga},
  \citenamefont {{van den Brink}},\ and\ \citenamefont {Facio}}]{Guo2023}%
  \BibitemOpen
  \bibfield  {author} {\bibinfo {author} {\bibfnamefont {Y.}~\bibnamefont
  {Guo}}, \bibinfo {author} {\bibfnamefont {H.}~\bibnamefont {Liu}}, \bibinfo
  {author} {\bibfnamefont {O.}~\bibnamefont {Janson}}, \bibinfo {author}
  {\bibfnamefont {I.~C.}\ \bibnamefont {Fulga}}, \bibinfo {author}
  {\bibfnamefont {J.}~\bibnamefont {{van den Brink}}},\ and\ \bibinfo {author}
  {\bibfnamefont {J.~I.}\ \bibnamefont {Facio}},\ }\bibfield  {title} {\bibinfo
  {title} {Spin-split collinear antiferromagnets: A large-scale ab-initio
  study},\ }\href
  {https://doi.org/https://doi.org/10.1016/j.mtphys.2023.100991} {\bibfield
  {journal} {\bibinfo  {journal} {Mater. Today Phys.}\ }\textbf {\bibinfo
  {volume} {32}},\ \bibinfo {pages} {100991} (\bibinfo {year}
  {2023})}\BibitemShut {NoStop}%
\bibitem [{\citenamefont {Zeng}\ and\ \citenamefont {Zhao}(2024)}]{Zeng2024}%
  \BibitemOpen
  \bibfield  {author} {\bibinfo {author} {\bibfnamefont {S.}~\bibnamefont
  {Zeng}}\ and\ \bibinfo {author} {\bibfnamefont {Y.-J.}\ \bibnamefont
  {Zhao}},\ }\bibfield  {title} {\bibinfo {title} {Description of
  two-dimensional altermagnetism: Categorization using spin group theory},\
  }\href {https://doi.org/10.1103/PhysRevB.110.054406} {\bibfield  {journal}
  {\bibinfo  {journal} {Phys. Rev. B}\ }\textbf {\bibinfo {volume} {110}},\
  \bibinfo {pages} {054406} (\bibinfo {year} {2024})}\BibitemShut {NoStop}%
\bibitem [{\citenamefont {Lee}\ \emph {et~al.}(2024{\natexlab{a}})\citenamefont
  {Lee}, \citenamefont {Lee}, \citenamefont {Jung}, \citenamefont {Jung},
  \citenamefont {Kim}, \citenamefont {Lee}, \citenamefont {Seok}, \citenamefont
  {Kim}, \citenamefont {Park}, \citenamefont {\ifmmode~\check{S}\else
  \v{S}\fi{}mejkal}, \citenamefont {Kang},\ and\ \citenamefont
  {Kim}}]{Lee2024PRL}%
  \BibitemOpen
  \bibfield  {author} {\bibinfo {author} {\bibfnamefont {S.}~\bibnamefont
  {Lee}}, \bibinfo {author} {\bibfnamefont {S.}~\bibnamefont {Lee}}, \bibinfo
  {author} {\bibfnamefont {S.}~\bibnamefont {Jung}}, \bibinfo {author}
  {\bibfnamefont {J.}~\bibnamefont {Jung}}, \bibinfo {author} {\bibfnamefont
  {D.}~\bibnamefont {Kim}}, \bibinfo {author} {\bibfnamefont {Y.}~\bibnamefont
  {Lee}}, \bibinfo {author} {\bibfnamefont {B.}~\bibnamefont {Seok}}, \bibinfo
  {author} {\bibfnamefont {J.}~\bibnamefont {Kim}}, \bibinfo {author}
  {\bibfnamefont {B.~G.}\ \bibnamefont {Park}}, \bibinfo {author}
  {\bibfnamefont {L.}~\bibnamefont {\ifmmode~\check{S}\else \v{S}\fi{}mejkal}},
  \bibinfo {author} {\bibfnamefont {C.-J.}\ \bibnamefont {Kang}},\ and\
  \bibinfo {author} {\bibfnamefont {C.}~\bibnamefont {Kim}},\ }\bibfield
  {title} {\bibinfo {title} {Broken {K}ramers degeneracy in altermagnetic
  {MnTe}},\ }\href {https://doi.org/10.1103/PhysRevLett.132.036702} {\bibfield
  {journal} {\bibinfo  {journal} {Phys. Rev. Lett.}\ }\textbf {\bibinfo
  {volume} {132}},\ \bibinfo {pages} {036702} (\bibinfo {year}
  {2024}{\natexlab{a}})}\BibitemShut {NoStop}%
\bibitem [{\citenamefont {Krempask{\'y}}\ \emph {et~al.}(2024)\citenamefont
  {Krempask{\'y}}, \citenamefont {{\v S}mejkal}, \citenamefont {D'Souza},
  \citenamefont {Hajlaoui}, \citenamefont {Springholz}, \citenamefont
  {Uhl{\'\i}{\v r}ov{\'a}}, \citenamefont {Alarab}, \citenamefont
  {Constantinou}, \citenamefont {Strocov}, \citenamefont {Usanov},
  \citenamefont {Pudelko}, \citenamefont {Gonz{\'a}lez-Hern{\'a}ndez},
  \citenamefont {Birk~Hellenes}, \citenamefont {Jansa}, \citenamefont
  {Reichlov{\'a}}, \citenamefont {{\v S}ob{\'a}{\v n}}, \citenamefont
  {Gonzalez~Betancourt}, \citenamefont {Wadley}, \citenamefont {Sinova},
  \citenamefont {Kriegner}, \citenamefont {Min{\'a}r}, \citenamefont {Dil},\
  and\ \citenamefont {Jungwirth}}]{Krempask2024}%
  \BibitemOpen
  \bibfield  {author} {\bibinfo {author} {\bibfnamefont {J.}~\bibnamefont
  {Krempask{\'y}}}, \bibinfo {author} {\bibfnamefont {L.}~\bibnamefont {{\v
  S}mejkal}}, \bibinfo {author} {\bibfnamefont {S.~W.}\ \bibnamefont
  {D'Souza}}, \bibinfo {author} {\bibfnamefont {M.}~\bibnamefont {Hajlaoui}},
  \bibinfo {author} {\bibfnamefont {G.}~\bibnamefont {Springholz}}, \bibinfo
  {author} {\bibfnamefont {K.}~\bibnamefont {Uhl{\'\i}{\v r}ov{\'a}}}, \bibinfo
  {author} {\bibfnamefont {F.}~\bibnamefont {Alarab}}, \bibinfo {author}
  {\bibfnamefont {P.~C.}\ \bibnamefont {Constantinou}}, \bibinfo {author}
  {\bibfnamefont {V.}~\bibnamefont {Strocov}}, \bibinfo {author} {\bibfnamefont
  {D.}~\bibnamefont {Usanov}}, \bibinfo {author} {\bibfnamefont {W.~R.}\
  \bibnamefont {Pudelko}}, \bibinfo {author} {\bibfnamefont {R.}~\bibnamefont
  {Gonz{\'a}lez-Hern{\'a}ndez}}, \bibinfo {author} {\bibfnamefont
  {A.}~\bibnamefont {Birk~Hellenes}}, \bibinfo {author} {\bibfnamefont
  {Z.}~\bibnamefont {Jansa}}, \bibinfo {author} {\bibfnamefont
  {H.}~\bibnamefont {Reichlov{\'a}}}, \bibinfo {author} {\bibfnamefont
  {Z.}~\bibnamefont {{\v S}ob{\'a}{\v n}}}, \bibinfo {author} {\bibfnamefont
  {R.~D.}\ \bibnamefont {Gonzalez~Betancourt}}, \bibinfo {author}
  {\bibfnamefont {P.}~\bibnamefont {Wadley}}, \bibinfo {author} {\bibfnamefont
  {J.}~\bibnamefont {Sinova}}, \bibinfo {author} {\bibfnamefont
  {D.}~\bibnamefont {Kriegner}}, \bibinfo {author} {\bibfnamefont
  {J.}~\bibnamefont {Min{\'a}r}}, \bibinfo {author} {\bibfnamefont {J.~H.}\
  \bibnamefont {Dil}},\ and\ \bibinfo {author} {\bibfnamefont {T.}~\bibnamefont
  {Jungwirth}},\ }\bibfield  {title} {\bibinfo {title} {Altermagnetic lifting
  of {K}ramers spin degeneracy},\ }\href
  {https://doi.org/10.1038/s41586-023-06907-7} {\bibfield  {journal} {\bibinfo
  {journal} {Nature}\ }\textbf {\bibinfo {volume} {626}},\ \bibinfo {pages}
  {517} (\bibinfo {year} {2024})}\BibitemShut {NoStop}%
\bibitem [{\citenamefont {Reimers}\ \emph {et~al.}(2024)\citenamefont
  {Reimers}, \citenamefont {Odenbreit}, \citenamefont {Šmejkal}, \citenamefont
  {Strocov}, \citenamefont {Constantinou}, \citenamefont {Hellenes},
  \citenamefont {Jaeschke~Ubiergo}, \citenamefont {Campos}, \citenamefont
  {Bharadwaj}, \citenamefont {Chakraborty}, \citenamefont {Denneulin},
  \citenamefont {Shi}, \citenamefont {Dunin-Borkowski}, \citenamefont {Das},
  \citenamefont {Kläui}, \citenamefont {Sinova},\ and\ \citenamefont
  {Jourdan}}]{Reimers2024}%
  \BibitemOpen
  \bibfield  {author} {\bibinfo {author} {\bibfnamefont {S.}~\bibnamefont
  {Reimers}}, \bibinfo {author} {\bibfnamefont {L.}~\bibnamefont {Odenbreit}},
  \bibinfo {author} {\bibfnamefont {L.}~\bibnamefont {Šmejkal}}, \bibinfo
  {author} {\bibfnamefont {V.~N.}\ \bibnamefont {Strocov}}, \bibinfo {author}
  {\bibfnamefont {P.}~\bibnamefont {Constantinou}}, \bibinfo {author}
  {\bibfnamefont {A.~B.}\ \bibnamefont {Hellenes}}, \bibinfo {author}
  {\bibfnamefont {R.}~\bibnamefont {Jaeschke~Ubiergo}}, \bibinfo {author}
  {\bibfnamefont {W.~H.}\ \bibnamefont {Campos}}, \bibinfo {author}
  {\bibfnamefont {V.~K.}\ \bibnamefont {Bharadwaj}}, \bibinfo {author}
  {\bibfnamefont {A.}~\bibnamefont {Chakraborty}}, \bibinfo {author}
  {\bibfnamefont {T.}~\bibnamefont {Denneulin}}, \bibinfo {author}
  {\bibfnamefont {W.}~\bibnamefont {Shi}}, \bibinfo {author} {\bibfnamefont
  {R.~E.}\ \bibnamefont {Dunin-Borkowski}}, \bibinfo {author} {\bibfnamefont
  {S.}~\bibnamefont {Das}}, \bibinfo {author} {\bibfnamefont {M.}~\bibnamefont
  {Kläui}}, \bibinfo {author} {\bibfnamefont {J.}~\bibnamefont {Sinova}},\
  and\ \bibinfo {author} {\bibfnamefont {M.}~\bibnamefont {Jourdan}},\
  }\bibfield  {title} {\bibinfo {title} {Direct observation of altermagnetic
  band splitting in {CrSb} thin films},\ }\href
  {https://doi.org/10.1038/s41467-024-46476-5} {\bibfield  {journal} {\bibinfo
  {journal} {Nat. Commun.}\ }\textbf {\bibinfo {volume} {15}},\ \bibinfo
  {pages} {2116} (\bibinfo {year} {2024})}\BibitemShut {NoStop}%
\bibitem [{\citenamefont {Aoyama}\ and\ \citenamefont
  {Ohgushi}(2024)}]{Aoyama2024}%
  \BibitemOpen
  \bibfield  {author} {\bibinfo {author} {\bibfnamefont {T.}~\bibnamefont
  {Aoyama}}\ and\ \bibinfo {author} {\bibfnamefont {K.}~\bibnamefont
  {Ohgushi}},\ }\bibfield  {title} {\bibinfo {title} {Piezomagnetic properties
  in altermagnetic \ce{MnTe}},\ }\href
  {https://doi.org/10.1103/PhysRevMaterials.8.L041402} {\bibfield  {journal}
  {\bibinfo  {journal} {Phys. Rev. Mater.}\ }\textbf {\bibinfo {volume} {8}},\
  \bibinfo {pages} {L041402} (\bibinfo {year} {2024})}\BibitemShut {NoStop}%
\bibitem [{\citenamefont {Lin}\ \emph {et~al.}(2024)\citenamefont {Lin},
  \citenamefont {Chen}, \citenamefont {Lu}, \citenamefont {Liang},
  \citenamefont {Feng}, \citenamefont {Yamagami}, \citenamefont {Osiecki},
  \citenamefont {Leandersson}, \citenamefont {Thiagarajan}, \citenamefont
  {Liu}, \citenamefont {Felser},\ and\ \citenamefont {Ma}}]{Lin2024}%
  \BibitemOpen
  \bibfield  {author} {\bibinfo {author} {\bibfnamefont {Z.}~\bibnamefont
  {Lin}}, \bibinfo {author} {\bibfnamefont {D.}~\bibnamefont {Chen}}, \bibinfo
  {author} {\bibfnamefont {W.}~\bibnamefont {Lu}}, \bibinfo {author}
  {\bibfnamefont {X.}~\bibnamefont {Liang}}, \bibinfo {author} {\bibfnamefont
  {S.}~\bibnamefont {Feng}}, \bibinfo {author} {\bibfnamefont {K.}~\bibnamefont
  {Yamagami}}, \bibinfo {author} {\bibfnamefont {J.}~\bibnamefont {Osiecki}},
  \bibinfo {author} {\bibfnamefont {M.}~\bibnamefont {Leandersson}}, \bibinfo
  {author} {\bibfnamefont {B.}~\bibnamefont {Thiagarajan}}, \bibinfo {author}
  {\bibfnamefont {J.}~\bibnamefont {Liu}}, \bibinfo {author} {\bibfnamefont
  {C.}~\bibnamefont {Felser}},\ and\ \bibinfo {author} {\bibfnamefont
  {J.}~\bibnamefont {Ma}},\ }\href@noop {} {\bibinfo {title} {Observation of
  giant spin splitting and d-wave spin texture in room temperature altermagnet
  {RuO$_2$}}} (\bibinfo {year} {2024}),\ \Eprint
  {https://arxiv.org/abs/2402.04995} {arXiv:2402.04995 [cond-mat.mtrl-sci]}
  \BibitemShut {NoStop}%
\bibitem [{\citenamefont {Ahn}\ \emph {et~al.}(2019)\citenamefont {Ahn},
  \citenamefont {Hariki}, \citenamefont {Lee},\ and\ \citenamefont
  {Kune\ifmmode~\check{s}\else \v{s}\fi{}}}]{Kyo-Hoon2019}%
  \BibitemOpen
  \bibfield  {author} {\bibinfo {author} {\bibfnamefont {K.-H.}\ \bibnamefont
  {Ahn}}, \bibinfo {author} {\bibfnamefont {A.}~\bibnamefont {Hariki}},
  \bibinfo {author} {\bibfnamefont {K.-W.}\ \bibnamefont {Lee}},\ and\ \bibinfo
  {author} {\bibfnamefont {J.}~\bibnamefont {Kune\ifmmode~\check{s}\else
  \v{s}\fi{}}},\ }\bibfield  {title} {\bibinfo {title} {Antiferromagnetism in
  {RuO}$_{2}$ as $d$-wave {Pomeranchuk} instability},\ }\href
  {https://doi.org/10.1103/PhysRevB.99.184432} {\bibfield  {journal} {\bibinfo
  {journal} {Phys. Rev. B}\ }\textbf {\bibinfo {volume} {99}},\ \bibinfo
  {pages} {184432} (\bibinfo {year} {2019})}\BibitemShut {NoStop}%
\bibitem [{\citenamefont {Šmejkal}\ \emph {et~al.}(2020)\citenamefont
  {Šmejkal}, \citenamefont {González-Hernández}, \citenamefont {Jungwirth},\
  and\ \citenamefont {Sinova}}]{Libor2020}%
  \BibitemOpen
  \bibfield  {author} {\bibinfo {author} {\bibfnamefont {L.}~\bibnamefont
  {Šmejkal}}, \bibinfo {author} {\bibfnamefont {R.}~\bibnamefont
  {González-Hernández}}, \bibinfo {author} {\bibfnamefont {T.}~\bibnamefont
  {Jungwirth}},\ and\ \bibinfo {author} {\bibfnamefont {J.}~\bibnamefont
  {Sinova}},\ }\bibfield  {title} {\bibinfo {title} {Crystal time-reversal
  symmetry breaking and spontaneous {H}all effect in collinear
  antiferromagnets},\ }\href {https://doi.org/10.1126/sciadv.aaz8809}
  {\bibfield  {journal} {\bibinfo  {journal} {Sci. Adv.}\ }\textbf {\bibinfo
  {volume} {6}},\ \bibinfo {pages} {eaaz8809} (\bibinfo {year} {2020})},\
  \Eprint
  {https://arxiv.org/abs/https://www.science.org/doi/pdf/10.1126/sciadv.aaz8809}
  {https://www.science.org/doi/pdf/10.1126/sciadv.aaz8809} \BibitemShut
  {NoStop}%
\bibitem [{\citenamefont {\ifmmode~\check{S}\else \v{S}\fi{}mejkal}\ \emph
  {et~al.}(2022{\natexlab{b}})\citenamefont {\ifmmode~\check{S}\else
  \v{S}\fi{}mejkal}, \citenamefont {Sinova},\ and\ \citenamefont
  {Jungwirth}}]{Libor2022Review}%
  \BibitemOpen
  \bibfield  {author} {\bibinfo {author} {\bibfnamefont {L.}~\bibnamefont
  {\ifmmode~\check{S}\else \v{S}\fi{}mejkal}}, \bibinfo {author} {\bibfnamefont
  {J.}~\bibnamefont {Sinova}},\ and\ \bibinfo {author} {\bibfnamefont
  {T.}~\bibnamefont {Jungwirth}},\ }\bibfield  {title} {\bibinfo {title}
  {Emerging research landscape of altermagnetism},\ }\href
  {https://doi.org/10.1103/PhysRevX.12.040501} {\bibfield  {journal} {\bibinfo
  {journal} {Phys. Rev. X}\ }\textbf {\bibinfo {volume} {12}},\ \bibinfo
  {pages} {040501} (\bibinfo {year} {2022}{\natexlab{b}})}\BibitemShut
  {NoStop}%
\bibitem [{\citenamefont {McClarty}\ and\ \citenamefont
  {Rau}(2024)}]{Paul2024}%
  \BibitemOpen
  \bibfield  {author} {\bibinfo {author} {\bibfnamefont {P.~A.}\ \bibnamefont
  {McClarty}}\ and\ \bibinfo {author} {\bibfnamefont {J.~G.}\ \bibnamefont
  {Rau}},\ }\bibfield  {title} {\bibinfo {title} {Landau theory of
  altermagnetism},\ }\href {https://doi.org/10.1103/PhysRevLett.132.176702}
  {\bibfield  {journal} {\bibinfo  {journal} {Phys. Rev. Lett.}\ }\textbf
  {\bibinfo {volume} {132}},\ \bibinfo {pages} {176702} (\bibinfo {year}
  {2024})}\BibitemShut {NoStop}%
\bibitem [{\citenamefont {Bai}\ \emph {et~al.}(2024)\citenamefont {Bai},
  \citenamefont {Feng}, \citenamefont {Liu}, \citenamefont {Šmejkal},
  \citenamefont {Mokrousov},\ and\ \citenamefont {Yao}}]{Bai2024}%
  \BibitemOpen
  \bibfield  {author} {\bibinfo {author} {\bibfnamefont {L.}~\bibnamefont
  {Bai}}, \bibinfo {author} {\bibfnamefont {W.}~\bibnamefont {Feng}}, \bibinfo
  {author} {\bibfnamefont {S.}~\bibnamefont {Liu}}, \bibinfo {author}
  {\bibfnamefont {L.}~\bibnamefont {Šmejkal}}, \bibinfo {author}
  {\bibfnamefont {Y.}~\bibnamefont {Mokrousov}},\ and\ \bibinfo {author}
  {\bibfnamefont {Y.}~\bibnamefont {Yao}},\ }\href
  {https://arxiv.org/abs/2406.02123} {\bibinfo {title} {Altermagnetism:
  Exploring new frontiers in magnetism and spintronics}} (\bibinfo {year}
  {2024}),\ \Eprint {https://arxiv.org/abs/2406.02123} {arXiv:2406.02123
  [cond-mat.mtrl-sci]} \BibitemShut {NoStop}%
\bibitem [{\citenamefont {Naka}\ \emph {et~al.}(2019)\citenamefont {Naka},
  \citenamefont {Hayami}, \citenamefont {Kusunose}, \citenamefont {Yanagi},
  \citenamefont {Motome},\ and\ \citenamefont {Seo}}]{Naka2019}%
  \BibitemOpen
  \bibfield  {author} {\bibinfo {author} {\bibfnamefont {M.}~\bibnamefont
  {Naka}}, \bibinfo {author} {\bibfnamefont {S.}~\bibnamefont {Hayami}},
  \bibinfo {author} {\bibfnamefont {H.}~\bibnamefont {Kusunose}}, \bibinfo
  {author} {\bibfnamefont {Y.}~\bibnamefont {Yanagi}}, \bibinfo {author}
  {\bibfnamefont {Y.}~\bibnamefont {Motome}},\ and\ \bibinfo {author}
  {\bibfnamefont {H.}~\bibnamefont {Seo}},\ }\bibfield  {title} {\bibinfo
  {title} {Spin current generation in organic antiferromagnets},\ }\href
  {https://doi.org/10.1038/s41467-019-12229-y} {\bibfield  {journal} {\bibinfo
  {journal} {Nat. Commun.}\ }\textbf {\bibinfo {volume} {10}},\ \bibinfo
  {pages} {4305} (\bibinfo {year} {2019})}\BibitemShut {NoStop}%
\bibitem [{\citenamefont {Gonz\'alez-Hern\'andez}\ \emph
  {et~al.}(2021)\citenamefont {Gonz\'alez-Hern\'andez}, \citenamefont
  {\ifmmode~\check{S}\else \v{S}\fi{}mejkal}, \citenamefont {V\'yborn\'y},
  \citenamefont {Yahagi}, \citenamefont {Sinova}, \citenamefont {Jungwirth},\
  and\ \citenamefont {\ifmmode~\check{Z}\else
  \v{Z}\fi{}elezn\'y}}]{Hernandez2021}%
  \BibitemOpen
  \bibfield  {author} {\bibinfo {author} {\bibfnamefont {R.}~\bibnamefont
  {Gonz\'alez-Hern\'andez}}, \bibinfo {author} {\bibfnamefont {L.}~\bibnamefont
  {\ifmmode~\check{S}\else \v{S}\fi{}mejkal}}, \bibinfo {author} {\bibfnamefont
  {K.}~\bibnamefont {V\'yborn\'y}}, \bibinfo {author} {\bibfnamefont
  {Y.}~\bibnamefont {Yahagi}}, \bibinfo {author} {\bibfnamefont
  {J.}~\bibnamefont {Sinova}}, \bibinfo {author} {\bibfnamefont
  {T.}~\bibnamefont {Jungwirth}},\ and\ \bibinfo {author} {\bibfnamefont
  {J.}~\bibnamefont {\ifmmode~\check{Z}\else \v{Z}\fi{}elezn\'y}},\ }\bibfield
  {title} {\bibinfo {title} {Efficient electrical spin splitter based on
  nonrelativistic collinear antiferromagnetism},\ }\href
  {https://doi.org/10.1103/PhysRevLett.126.127701} {\bibfield  {journal}
  {\bibinfo  {journal} {Phys. Rev. Lett.}\ }\textbf {\bibinfo {volume} {126}},\
  \bibinfo {pages} {127701} (\bibinfo {year} {2021})}\BibitemShut {NoStop}%
\bibitem [{\citenamefont {Shao}\ \emph {et~al.}(2021)\citenamefont {Shao},
  \citenamefont {Zhang}, \citenamefont {Li}, \citenamefont {Eom},\ and\
  \citenamefont {Tsymbal}}]{Shao2021}%
  \BibitemOpen
  \bibfield  {author} {\bibinfo {author} {\bibfnamefont {D.-F.}\ \bibnamefont
  {Shao}}, \bibinfo {author} {\bibfnamefont {S.-H.}\ \bibnamefont {Zhang}},
  \bibinfo {author} {\bibfnamefont {M.}~\bibnamefont {Li}}, \bibinfo {author}
  {\bibfnamefont {C.-B.}\ \bibnamefont {Eom}},\ and\ \bibinfo {author}
  {\bibfnamefont {E.~Y.}\ \bibnamefont {Tsymbal}},\ }\bibfield  {title}
  {\bibinfo {title} {Spin-neutral currents for spintronics},\ }\href
  {https://doi.org/10.1038/s41467-021-26915-3} {\bibfield  {journal} {\bibinfo
  {journal} {Nat. Commun.}\ }\textbf {\bibinfo {volume} {12}},\ \bibinfo
  {pages} {7061} (\bibinfo {year} {2021})}\BibitemShut {NoStop}%
\bibitem [{\citenamefont {Bose}\ \emph {et~al.}(2022)\citenamefont {Bose},
  \citenamefont {Schreiber}, \citenamefont {Jain}, \citenamefont {Shao},
  \citenamefont {Nair}, \citenamefont {Sun}, \citenamefont {Zhang},
  \citenamefont {Muller}, \citenamefont {Tsymbal}, \citenamefont {Schlom},\
  and\ \citenamefont {Ralph}}]{Bose2022}%
  \BibitemOpen
  \bibfield  {author} {\bibinfo {author} {\bibfnamefont {A.}~\bibnamefont
  {Bose}}, \bibinfo {author} {\bibfnamefont {N.~J.}\ \bibnamefont {Schreiber}},
  \bibinfo {author} {\bibfnamefont {R.}~\bibnamefont {Jain}}, \bibinfo {author}
  {\bibfnamefont {D.-F.}\ \bibnamefont {Shao}}, \bibinfo {author}
  {\bibfnamefont {H.~P.}\ \bibnamefont {Nair}}, \bibinfo {author}
  {\bibfnamefont {J.}~\bibnamefont {Sun}}, \bibinfo {author} {\bibfnamefont
  {X.~S.}\ \bibnamefont {Zhang}}, \bibinfo {author} {\bibfnamefont {D.~A.}\
  \bibnamefont {Muller}}, \bibinfo {author} {\bibfnamefont {E.~Y.}\
  \bibnamefont {Tsymbal}}, \bibinfo {author} {\bibfnamefont {D.~G.}\
  \bibnamefont {Schlom}},\ and\ \bibinfo {author} {\bibfnamefont {D.~C.}\
  \bibnamefont {Ralph}},\ }\bibfield  {title} {\bibinfo {title} {Tilted spin
  current generated by the collinear antiferromagnet ruthenium dioxide},\
  }\href {https://doi.org/10.1038/s41928-022-00744-8} {\bibfield  {journal}
  {\bibinfo  {journal} {Nat. Electron.}\ }\textbf {\bibinfo {volume} {5}},\
  \bibinfo {pages} {267} (\bibinfo {year} {2022})}\BibitemShut {NoStop}%
\bibitem [{\citenamefont {Hu}\ \emph {et~al.}(2024)\citenamefont {Hu},
  \citenamefont {Janson}, \citenamefont {Felser}, \citenamefont {McClarty},
  \citenamefont {van~den Brink},\ and\ \citenamefont {Vergniory}}]{Hu2024}%
  \BibitemOpen
  \bibfield  {author} {\bibinfo {author} {\bibfnamefont {M.}~\bibnamefont
  {Hu}}, \bibinfo {author} {\bibfnamefont {O.}~\bibnamefont {Janson}}, \bibinfo
  {author} {\bibfnamefont {C.}~\bibnamefont {Felser}}, \bibinfo {author}
  {\bibfnamefont {P.}~\bibnamefont {McClarty}}, \bibinfo {author}
  {\bibfnamefont {J.}~\bibnamefont {van~den Brink}},\ and\ \bibinfo {author}
  {\bibfnamefont {M.~G.}\ \bibnamefont {Vergniory}},\ }\href
  {https://arxiv.org/abs/2410.17993} {\bibinfo {title} {Spin {H}all and
  edelstein effects in novel chiral noncollinear altermagnets}} (\bibinfo
  {year} {2024}),\ \Eprint {https://arxiv.org/abs/2410.17993} {arXiv:2410.17993
  [cond-mat.mtrl-sci]} \BibitemShut {NoStop}%
\bibitem [{\citenamefont {Bai}\ \emph {et~al.}(2022)\citenamefont {Bai},
  \citenamefont {Han}, \citenamefont {Feng}, \citenamefont {Zhou},
  \citenamefont {Su}, \citenamefont {Wang}, \citenamefont {Liao}, \citenamefont
  {Zhu}, \citenamefont {Chen}, \citenamefont {Pan}, \citenamefont {Fan},\ and\
  \citenamefont {Song}}]{Bai2022}%
  \BibitemOpen
  \bibfield  {author} {\bibinfo {author} {\bibfnamefont {H.}~\bibnamefont
  {Bai}}, \bibinfo {author} {\bibfnamefont {L.}~\bibnamefont {Han}}, \bibinfo
  {author} {\bibfnamefont {X.~Y.}\ \bibnamefont {Feng}}, \bibinfo {author}
  {\bibfnamefont {Y.~J.}\ \bibnamefont {Zhou}}, \bibinfo {author}
  {\bibfnamefont {R.~X.}\ \bibnamefont {Su}}, \bibinfo {author} {\bibfnamefont
  {Q.}~\bibnamefont {Wang}}, \bibinfo {author} {\bibfnamefont {L.~Y.}\
  \bibnamefont {Liao}}, \bibinfo {author} {\bibfnamefont {W.~X.}\ \bibnamefont
  {Zhu}}, \bibinfo {author} {\bibfnamefont {X.~Z.}\ \bibnamefont {Chen}},
  \bibinfo {author} {\bibfnamefont {F.}~\bibnamefont {Pan}}, \bibinfo {author}
  {\bibfnamefont {X.~L.}\ \bibnamefont {Fan}},\ and\ \bibinfo {author}
  {\bibfnamefont {C.}~\bibnamefont {Song}},\ }\bibfield  {title} {\bibinfo
  {title} {Observation of spin splitting torque in a collinear antiferromagnet
  {RuO}$_{2}$},\ }\href {https://doi.org/10.1103/PhysRevLett.128.197202}
  {\bibfield  {journal} {\bibinfo  {journal} {Phys. Rev. Lett.}\ }\textbf
  {\bibinfo {volume} {128}},\ \bibinfo {pages} {197202} (\bibinfo {year}
  {2022})}\BibitemShut {NoStop}%
\bibitem [{\citenamefont {Karube}\ \emph {et~al.}(2022)\citenamefont {Karube},
  \citenamefont {Tanaka}, \citenamefont {Sugawara}, \citenamefont {Kadoguchi},
  \citenamefont {Kohda},\ and\ \citenamefont {Nitta}}]{Karube2021}%
  \BibitemOpen
  \bibfield  {author} {\bibinfo {author} {\bibfnamefont {S.}~\bibnamefont
  {Karube}}, \bibinfo {author} {\bibfnamefont {T.}~\bibnamefont {Tanaka}},
  \bibinfo {author} {\bibfnamefont {D.}~\bibnamefont {Sugawara}}, \bibinfo
  {author} {\bibfnamefont {N.}~\bibnamefont {Kadoguchi}}, \bibinfo {author}
  {\bibfnamefont {M.}~\bibnamefont {Kohda}},\ and\ \bibinfo {author}
  {\bibfnamefont {J.}~\bibnamefont {Nitta}},\ }\bibfield  {title} {\bibinfo
  {title} {Observation of spin-splitter torque in collinear antiferromagnetic
  {RuO}$_{2}$},\ }\href {https://doi.org/10.1103/PhysRevLett.129.137201}
  {\bibfield  {journal} {\bibinfo  {journal} {Phys. Rev. Lett.}\ }\textbf
  {\bibinfo {volume} {129}},\ \bibinfo {pages} {137201} (\bibinfo {year}
  {2022})}\BibitemShut {NoStop}%
\bibitem [{\citenamefont {\ifmmode~\check{S}\else \v{S}\fi{}mejkal}\ \emph
  {et~al.}(2022{\natexlab{c}})\citenamefont {\ifmmode~\check{S}\else
  \v{S}\fi{}mejkal}, \citenamefont {Hellenes}, \citenamefont
  {Gonz\'alez-Hern\'andez}, \citenamefont {Sinova},\ and\ \citenamefont
  {Jungwirth}}]{Libor2022}%
  \BibitemOpen
  \bibfield  {author} {\bibinfo {author} {\bibfnamefont {L.}~\bibnamefont
  {\ifmmode~\check{S}\else \v{S}\fi{}mejkal}}, \bibinfo {author} {\bibfnamefont
  {A.~B.}\ \bibnamefont {Hellenes}}, \bibinfo {author} {\bibfnamefont
  {R.}~\bibnamefont {Gonz\'alez-Hern\'andez}}, \bibinfo {author} {\bibfnamefont
  {J.}~\bibnamefont {Sinova}},\ and\ \bibinfo {author} {\bibfnamefont
  {T.}~\bibnamefont {Jungwirth}},\ }\bibfield  {title} {\bibinfo {title} {Giant
  and tunneling magnetoresistance in unconventional collinear antiferromagnets
  with nonrelativistic spin-momentum coupling},\ }\href
  {https://doi.org/10.1103/PhysRevX.12.011028} {\bibfield  {journal} {\bibinfo
  {journal} {Phys. Rev. X}\ }\textbf {\bibinfo {volume} {12}},\ \bibinfo
  {pages} {011028} (\bibinfo {year} {2022}{\natexlab{c}})}\BibitemShut
  {NoStop}%
\bibitem [{\citenamefont {{\v S}mejkal}\ \emph
  {et~al.}(2022{\natexlab{a}})\citenamefont {{\v S}mejkal}, \citenamefont
  {MacDonald}, \citenamefont {Sinova}, \citenamefont {Nakatsuji},\ and\
  \citenamefont {Jungwirth}}]{Libor2022NatRev}%
  \BibitemOpen
  \bibfield  {author} {\bibinfo {author} {\bibfnamefont {L.}~\bibnamefont {{\v
  S}mejkal}}, \bibinfo {author} {\bibfnamefont {A.~H.}\ \bibnamefont
  {MacDonald}}, \bibinfo {author} {\bibfnamefont {J.}~\bibnamefont {Sinova}},
  \bibinfo {author} {\bibfnamefont {S.}~\bibnamefont {Nakatsuji}},\ and\
  \bibinfo {author} {\bibfnamefont {T.}~\bibnamefont {Jungwirth}},\ }\bibfield
  {title} {\bibinfo {title} {Anomalous {H}all antiferromagnets},\ }\href
  {https://doi.org/10.1038/s41578-022-00430-3} {\bibfield  {journal} {\bibinfo
  {journal} {Nature Reviews Materials}\ }\textbf {\bibinfo {volume} {7}},\
  \bibinfo {pages} {482} (\bibinfo {year} {2022}{\natexlab{a}})}\BibitemShut
  {NoStop}%
\bibitem [{\citenamefont {Reichlová}\ \emph {et~al.}(2020)\citenamefont
  {Reichlová}, \citenamefont {Seeger}, \citenamefont {González-Hernández},
  \citenamefont {Kounta}, \citenamefont {Schlitz}, \citenamefont {Kriegner},
  \citenamefont {Ritzinger}, \citenamefont {Lammel}, \citenamefont {Leiviskä},
  \citenamefont {Petříček}, \citenamefont {Doležal}, \citenamefont
  {Schmoranzerová}, \citenamefont {Bad'ura}, \citenamefont {Thomas},
  \citenamefont {Baltz}, \citenamefont {Michez}, \citenamefont {Sinova},
  \citenamefont {Goennenwein}, \citenamefont {Jungwirth},\ and\ \citenamefont
  {Šmejkal}}]{Helena2020}%
  \BibitemOpen
  \bibfield  {author} {\bibinfo {author} {\bibfnamefont {H.}~\bibnamefont
  {Reichlová}}, \bibinfo {author} {\bibfnamefont {R.~L.}\ \bibnamefont
  {Seeger}}, \bibinfo {author} {\bibfnamefont {R.}~\bibnamefont
  {González-Hernández}}, \bibinfo {author} {\bibfnamefont {I.}~\bibnamefont
  {Kounta}}, \bibinfo {author} {\bibfnamefont {R.}~\bibnamefont {Schlitz}},
  \bibinfo {author} {\bibfnamefont {D.}~\bibnamefont {Kriegner}}, \bibinfo
  {author} {\bibfnamefont {P.}~\bibnamefont {Ritzinger}}, \bibinfo {author}
  {\bibfnamefont {M.}~\bibnamefont {Lammel}}, \bibinfo {author} {\bibfnamefont
  {M.}~\bibnamefont {Leiviskä}}, \bibinfo {author} {\bibfnamefont
  {V.}~\bibnamefont {Petříček}}, \bibinfo {author} {\bibfnamefont
  {P.}~\bibnamefont {Doležal}}, \bibinfo {author} {\bibfnamefont
  {E.}~\bibnamefont {Schmoranzerová}}, \bibinfo {author} {\bibfnamefont
  {A.}~\bibnamefont {Bad'ura}}, \bibinfo {author} {\bibfnamefont
  {A.}~\bibnamefont {Thomas}}, \bibinfo {author} {\bibfnamefont
  {V.}~\bibnamefont {Baltz}}, \bibinfo {author} {\bibfnamefont
  {L.}~\bibnamefont {Michez}}, \bibinfo {author} {\bibfnamefont
  {J.}~\bibnamefont {Sinova}}, \bibinfo {author} {\bibfnamefont {S.~T.~B.}\
  \bibnamefont {Goennenwein}}, \bibinfo {author} {\bibfnamefont
  {T.}~\bibnamefont {Jungwirth}},\ and\ \bibinfo {author} {\bibfnamefont
  {L.}~\bibnamefont {Šmejkal}},\ }\bibfield  {title} {\bibinfo {title}
  {Macroscopic time reversal symmetry breaking by staggered spin-momentum
  interaction},\ }\href {https://arxiv.org/abs/2012.15651} {\bibfield
  {journal} {\bibinfo  {journal} {arXiv 2012.15651}\ } (\bibinfo {year}
  {2020})}\BibitemShut {NoStop}%
\bibitem [{\citenamefont {Feng}\ \emph {et~al.}(2022)\citenamefont {Feng},
  \citenamefont {Zhou}, \citenamefont {{\v S}mejkal}, \citenamefont {Wu},
  \citenamefont {Zhu}, \citenamefont {Guo}, \citenamefont
  {Gonz{\'a}lez-Hern{\'a}ndez}, \citenamefont {Wang}, \citenamefont {Yan},
  \citenamefont {Qin}, \citenamefont {Zhang}, \citenamefont {Wu}, \citenamefont
  {Chen}, \citenamefont {Meng}, \citenamefont {Liu}, \citenamefont {Xia},
  \citenamefont {Sinova}, \citenamefont {Jungwirth},\ and\ \citenamefont
  {Liu}}]{Feng2020}%
  \BibitemOpen
  \bibfield  {author} {\bibinfo {author} {\bibfnamefont {Z.}~\bibnamefont
  {Feng}}, \bibinfo {author} {\bibfnamefont {X.}~\bibnamefont {Zhou}}, \bibinfo
  {author} {\bibfnamefont {L.}~\bibnamefont {{\v S}mejkal}}, \bibinfo {author}
  {\bibfnamefont {L.}~\bibnamefont {Wu}}, \bibinfo {author} {\bibfnamefont
  {Z.}~\bibnamefont {Zhu}}, \bibinfo {author} {\bibfnamefont {H.}~\bibnamefont
  {Guo}}, \bibinfo {author} {\bibfnamefont {R.}~\bibnamefont
  {Gonz{\'a}lez-Hern{\'a}ndez}}, \bibinfo {author} {\bibfnamefont
  {X.}~\bibnamefont {Wang}}, \bibinfo {author} {\bibfnamefont {H.}~\bibnamefont
  {Yan}}, \bibinfo {author} {\bibfnamefont {P.}~\bibnamefont {Qin}}, \bibinfo
  {author} {\bibfnamefont {X.}~\bibnamefont {Zhang}}, \bibinfo {author}
  {\bibfnamefont {H.}~\bibnamefont {Wu}}, \bibinfo {author} {\bibfnamefont
  {H.}~\bibnamefont {Chen}}, \bibinfo {author} {\bibfnamefont {Z.}~\bibnamefont
  {Meng}}, \bibinfo {author} {\bibfnamefont {L.}~\bibnamefont {Liu}}, \bibinfo
  {author} {\bibfnamefont {Z.}~\bibnamefont {Xia}}, \bibinfo {author}
  {\bibfnamefont {J.}~\bibnamefont {Sinova}}, \bibinfo {author} {\bibfnamefont
  {T.}~\bibnamefont {Jungwirth}},\ and\ \bibinfo {author} {\bibfnamefont
  {Z.}~\bibnamefont {Liu}},\ }\bibfield  {title} {\bibinfo {title} {An
  anomalous {H}all effect in altermagnetic ruthenium dioxide},\ }\href
  {https://doi.org/10.1038/s41928-022-00866-z} {\bibfield  {journal} {\bibinfo
  {journal} {Nat. Electron.}\ }\textbf {\bibinfo {volume} {5}},\ \bibinfo
  {pages} {735} (\bibinfo {year} {2022})}\BibitemShut {NoStop}%
\bibitem [{\citenamefont {Gonzalez~Betancourt}\ \emph
  {et~al.}(2023)\citenamefont {Gonzalez~Betancourt}, \citenamefont
  {Zub\'a\ifmmode~\check{c}\else \v{c}\fi{}}, \citenamefont
  {Gonzalez-Hernandez}, \citenamefont {Geishendorf}, \citenamefont {\ifmmode
  \check{S}\else \v{S}\fi{}ob\'a\ifmmode~\check{n}\else \v{n}\fi{}},
  \citenamefont {Springholz}, \citenamefont {Olejn\'{\i}k}, \citenamefont
  {\ifmmode~\check{S}\else \v{S}\fi{}mejkal}, \citenamefont {Sinova},
  \citenamefont {Jungwirth}, \citenamefont {Goennenwein}, \citenamefont
  {Thomas}, \citenamefont {Reichlov\'a}, \citenamefont {\ifmmode~\check{Z}\else
  \v{Z}\fi{}elezn\'y},\ and\ \citenamefont {Kriegner}}]{Betancourt2021}%
  \BibitemOpen
  \bibfield  {author} {\bibinfo {author} {\bibfnamefont {R.~D.}\ \bibnamefont
  {Gonzalez~Betancourt}}, \bibinfo {author} {\bibfnamefont {J.}~\bibnamefont
  {Zub\'a\ifmmode~\check{c}\else \v{c}\fi{}}}, \bibinfo {author} {\bibfnamefont
  {R.}~\bibnamefont {Gonzalez-Hernandez}}, \bibinfo {author} {\bibfnamefont
  {K.}~\bibnamefont {Geishendorf}}, \bibinfo {author} {\bibfnamefont
  {Z.}~\bibnamefont {\ifmmode \check{S}\else
  \v{S}\fi{}ob\'a\ifmmode~\check{n}\else \v{n}\fi{}}}, \bibinfo {author}
  {\bibfnamefont {G.}~\bibnamefont {Springholz}}, \bibinfo {author}
  {\bibfnamefont {K.}~\bibnamefont {Olejn\'{\i}k}}, \bibinfo {author}
  {\bibfnamefont {L.}~\bibnamefont {\ifmmode~\check{S}\else \v{S}\fi{}mejkal}},
  \bibinfo {author} {\bibfnamefont {J.}~\bibnamefont {Sinova}}, \bibinfo
  {author} {\bibfnamefont {T.}~\bibnamefont {Jungwirth}}, \bibinfo {author}
  {\bibfnamefont {S.~T.~B.}\ \bibnamefont {Goennenwein}}, \bibinfo {author}
  {\bibfnamefont {A.}~\bibnamefont {Thomas}}, \bibinfo {author} {\bibfnamefont
  {H.}~\bibnamefont {Reichlov\'a}}, \bibinfo {author} {\bibfnamefont
  {J.}~\bibnamefont {\ifmmode~\check{Z}\else \v{Z}\fi{}elezn\'y}},\ and\
  \bibinfo {author} {\bibfnamefont {D.}~\bibnamefont {Kriegner}},\ }\bibfield
  {title} {\bibinfo {title} {Spontaneous anomalous {H}all effect arising from
  an unconventional compensated magnetic phase in a semiconductor},\ }\href
  {https://doi.org/10.1103/PhysRevLett.130.036702} {\bibfield  {journal}
  {\bibinfo  {journal} {Phys. Rev. Lett.}\ }\textbf {\bibinfo {volume} {130}},\
  \bibinfo {pages} {036702} (\bibinfo {year} {2023})}\BibitemShut {NoStop}%
\bibitem [{\citenamefont {{\v S}mejkal}\ \emph
  {et~al.}(2022{\natexlab{b}})\citenamefont {{\v S}mejkal}, \citenamefont
  {MacDonald}, \citenamefont {Sinova}, \citenamefont {Nakatsuji},\ and\
  \citenamefont {Jungwirth}}]{SmejkalAHE2022}%
  \BibitemOpen
  \bibfield  {author} {\bibinfo {author} {\bibfnamefont {L.}~\bibnamefont {{\v
  S}mejkal}}, \bibinfo {author} {\bibfnamefont {A.~H.}\ \bibnamefont
  {MacDonald}}, \bibinfo {author} {\bibfnamefont {J.}~\bibnamefont {Sinova}},
  \bibinfo {author} {\bibfnamefont {S.}~\bibnamefont {Nakatsuji}},\ and\
  \bibinfo {author} {\bibfnamefont {T.}~\bibnamefont {Jungwirth}},\ }\bibfield
  {title} {\bibinfo {title} {Anomalous {H}all antiferromagnets},\ }\href
  {https://doi.org/10.1038/s41578-022-00430-3} {\bibfield  {journal} {\bibinfo
  {journal} {Nat. Rev. Mater.}\ }\textbf {\bibinfo {volume} {7}},\ \bibinfo
  {pages} {482} (\bibinfo {year} {2022}{\natexlab{b}})}\BibitemShut {NoStop}%
\bibitem [{\citenamefont {Cheong}\ and\ \citenamefont
  {Huang}(2024)}]{Cheong2024}%
  \BibitemOpen
  \bibfield  {author} {\bibinfo {author} {\bibfnamefont {S.-W.}\ \bibnamefont
  {Cheong}}\ and\ \bibinfo {author} {\bibfnamefont {F.-T.}\ \bibnamefont
  {Huang}},\ }\bibfield  {title} {\bibinfo {title} {Altermagnetism with
  non-collinear spins},\ }\href {https://doi.org/10.1038/s41535-024-00626-6}
  {\bibfield  {journal} {\bibinfo  {journal} {npj Quantum Materials}\ }\textbf
  {\bibinfo {volume} {9}},\ \bibinfo {pages} {13} (\bibinfo {year}
  {2024})}\BibitemShut {NoStop}%
\bibitem [{\citenamefont {Sato}\ \emph {et~al.}(2024)\citenamefont {Sato},
  \citenamefont {Haddad}, \citenamefont {Fulga}, \citenamefont {Assaad},\ and\
  \citenamefont {van~den Brink}}]{Sato2024}%
  \BibitemOpen
  \bibfield  {author} {\bibinfo {author} {\bibfnamefont {T.}~\bibnamefont
  {Sato}}, \bibinfo {author} {\bibfnamefont {S.}~\bibnamefont {Haddad}},
  \bibinfo {author} {\bibfnamefont {I.~C.}\ \bibnamefont {Fulga}}, \bibinfo
  {author} {\bibfnamefont {F.~F.}\ \bibnamefont {Assaad}},\ and\ \bibinfo
  {author} {\bibfnamefont {J.}~\bibnamefont {van~den Brink}},\ }\bibfield
  {title} {\bibinfo {title} {Altermagnetic anomalous {H}all effect emerging
  from electronic correlations},\ }\href
  {https://doi.org/10.1103/PhysRevLett.133.086503} {\bibfield  {journal}
  {\bibinfo  {journal} {Phys. Rev. Lett.}\ }\textbf {\bibinfo {volume} {133}},\
  \bibinfo {pages} {086503} (\bibinfo {year} {2024})}\BibitemShut {NoStop}%
\bibitem [{\citenamefont {Mazin}(2022)}]{Mazin2022}%
  \BibitemOpen
  \bibfield  {author} {\bibinfo {author} {\bibfnamefont {I.~I.}\ \bibnamefont
  {Mazin}},\ }\bibfield  {title} {\bibinfo {title} {Notes on altermagnetism and
  superconductivity},\ }\href {https://arxiv.org/abs/2203.05000} {\bibfield
  {journal} {\bibinfo  {journal} {arXiv 2203.05000}\ } (\bibinfo {year}
  {2022})}\BibitemShut {NoStop}%
\bibitem [{\citenamefont {Zhu}\ \emph {et~al.}(2023)\citenamefont {Zhu},
  \citenamefont {Zhuang}, \citenamefont {Wu},\ and\ \citenamefont
  {Yan}}]{Zhu2023}%
  \BibitemOpen
  \bibfield  {author} {\bibinfo {author} {\bibfnamefont {D.}~\bibnamefont
  {Zhu}}, \bibinfo {author} {\bibfnamefont {Z.-Y.}\ \bibnamefont {Zhuang}},
  \bibinfo {author} {\bibfnamefont {Z.}~\bibnamefont {Wu}},\ and\ \bibinfo
  {author} {\bibfnamefont {Z.}~\bibnamefont {Yan}},\ }\bibfield  {title}
  {\bibinfo {title} {Topological superconductivity in two-dimensional
  altermagnetic metals},\ }\href {https://doi.org/10.1103/PhysRevB.108.184505}
  {\bibfield  {journal} {\bibinfo  {journal} {Phys. Rev. B}\ }\textbf {\bibinfo
  {volume} {108}},\ \bibinfo {pages} {184505} (\bibinfo {year}
  {2023})}\BibitemShut {NoStop}%
\bibitem [{\citenamefont {Banerjee}\ and\ \citenamefont
  {Scheurer}(2024)}]{Banerjee2024}%
  \BibitemOpen
  \bibfield  {author} {\bibinfo {author} {\bibfnamefont {S.}~\bibnamefont
  {Banerjee}}\ and\ \bibinfo {author} {\bibfnamefont {M.~S.}\ \bibnamefont
  {Scheurer}},\ }\bibfield  {title} {\bibinfo {title} {Altermagnetic
  superconducting diode effect},\ }\href
  {https://doi.org/10.1103/PhysRevB.110.024503} {\bibfield  {journal} {\bibinfo
   {journal} {Phys. Rev. B}\ }\textbf {\bibinfo {volume} {110}},\ \bibinfo
  {pages} {024503} (\bibinfo {year} {2024})}\BibitemShut {NoStop}%
\bibitem [{\citenamefont {Chakraborty}\ and\ \citenamefont
  {Black-Schaffer}(2024)}]{Chakraborty2024}%
  \BibitemOpen
  \bibfield  {author} {\bibinfo {author} {\bibfnamefont {D.}~\bibnamefont
  {Chakraborty}}\ and\ \bibinfo {author} {\bibfnamefont {A.~M.}\ \bibnamefont
  {Black-Schaffer}},\ }\bibfield  {title} {\bibinfo {title} {Zero-field
  finite-momentum and field-induced superconductivity in altermagnets},\ }\href
  {https://doi.org/10.1103/PhysRevB.110.L060508} {\bibfield  {journal}
  {\bibinfo  {journal} {Phys. Rev. B}\ }\textbf {\bibinfo {volume} {110}},\
  \bibinfo {pages} {L060508} (\bibinfo {year} {2024})}\BibitemShut {NoStop}%
\bibitem [{\citenamefont {Zhang}\ \emph {et~al.}(2024)\citenamefont {Zhang},
  \citenamefont {Hu},\ and\ \citenamefont {Neupert}}]{Zhang2024}%
  \BibitemOpen
  \bibfield  {author} {\bibinfo {author} {\bibfnamefont {S.-B.}\ \bibnamefont
  {Zhang}}, \bibinfo {author} {\bibfnamefont {L.-H.}\ \bibnamefont {Hu}},\ and\
  \bibinfo {author} {\bibfnamefont {T.}~\bibnamefont {Neupert}},\ }\bibfield
  {title} {\bibinfo {title} {Finite-momentum cooper pairing in proximitized
  altermagnets},\ }\href {https://doi.org/10.1038/s41467-024-45951-3}
  {\bibfield  {journal} {\bibinfo  {journal} {Nat. Commun.}\ }\textbf {\bibinfo
  {volume} {15}},\ \bibinfo {pages} {1801} (\bibinfo {year}
  {2024})}\BibitemShut {NoStop}%
\bibitem [{\citenamefont {Lee}\ \emph {et~al.}(2024{\natexlab{b}})\citenamefont
  {Lee}, \citenamefont {Qian},\ and\ \citenamefont {Yang}}]{Lee2024}%
  \BibitemOpen
  \bibfield  {author} {\bibinfo {author} {\bibfnamefont {S.~H.}\ \bibnamefont
  {Lee}}, \bibinfo {author} {\bibfnamefont {Y.}~\bibnamefont {Qian}},\ and\
  \bibinfo {author} {\bibfnamefont {B.-J.}\ \bibnamefont {Yang}},\ }\bibfield
  {title} {\bibinfo {title} {Fermi surface spin texture and topological
  superconductivity in spin-orbit free noncollinear antiferromagnets},\ }\href
  {https://doi.org/10.1103/PhysRevLett.132.196602} {\bibfield  {journal}
  {\bibinfo  {journal} {Phys. Rev. Lett.}\ }\textbf {\bibinfo {volume} {132}},\
  \bibinfo {pages} {196602} (\bibinfo {year} {2024}{\natexlab{b}})}\BibitemShut
  {NoStop}%
\bibitem [{\citenamefont {Zhou}\ \emph {et~al.}(2024)\citenamefont {Zhou},
  \citenamefont {Feng}, \citenamefont {Zhang}, \citenamefont
  {\ifmmode~\check{S}\else \v{S}\fi{}mejkal}, \citenamefont {Sinova},
  \citenamefont {Mokrousov},\ and\ \citenamefont {Yao}}]{Zhou2024}%
  \BibitemOpen
  \bibfield  {author} {\bibinfo {author} {\bibfnamefont {X.}~\bibnamefont
  {Zhou}}, \bibinfo {author} {\bibfnamefont {W.}~\bibnamefont {Feng}}, \bibinfo
  {author} {\bibfnamefont {R.-W.}\ \bibnamefont {Zhang}}, \bibinfo {author}
  {\bibfnamefont {L.}~\bibnamefont {\ifmmode~\check{S}\else \v{S}\fi{}mejkal}},
  \bibinfo {author} {\bibfnamefont {J.}~\bibnamefont {Sinova}}, \bibinfo
  {author} {\bibfnamefont {Y.}~\bibnamefont {Mokrousov}},\ and\ \bibinfo
  {author} {\bibfnamefont {Y.}~\bibnamefont {Yao}},\ }\bibfield  {title}
  {\bibinfo {title} {Crystal thermal transport in altermagnetic
  ${\mathrm{ruo}}_{2}$},\ }\href
  {https://doi.org/10.1103/PhysRevLett.132.056701} {\bibfield  {journal}
  {\bibinfo  {journal} {Phys. Rev. Lett.}\ }\textbf {\bibinfo {volume} {132}},\
  \bibinfo {pages} {056701} (\bibinfo {year} {2024})}\BibitemShut {NoStop}%
\bibitem [{\citenamefont {Yershov}\ \emph {et~al.}(2024)\citenamefont
  {Yershov}, \citenamefont {Kravchuk}, \citenamefont {Daghofer},\ and\
  \citenamefont {van~den Brink}}]{Yershov2024}%
  \BibitemOpen
  \bibfield  {author} {\bibinfo {author} {\bibfnamefont {K.~V.}\ \bibnamefont
  {Yershov}}, \bibinfo {author} {\bibfnamefont {V.~P.}\ \bibnamefont
  {Kravchuk}}, \bibinfo {author} {\bibfnamefont {M.}~\bibnamefont {Daghofer}},\
  and\ \bibinfo {author} {\bibfnamefont {J.}~\bibnamefont {van~den Brink}},\
  }\bibfield  {title} {\bibinfo {title} {Fluctuation-induced piezomagnetism in
  local moment altermagnets},\ }\href
  {https://doi.org/10.1103/PhysRevB.110.144421} {\bibfield  {journal} {\bibinfo
   {journal} {Phys. Rev. B}\ }\textbf {\bibinfo {volume} {110}},\ \bibinfo
  {pages} {144421} (\bibinfo {year} {2024})}\BibitemShut {NoStop}%
\bibitem [{\citenamefont {\ifmmode~\check{S}\else \v{S}\fi{}mejkal}\ \emph
  {et~al.}(2023)\citenamefont {\ifmmode~\check{S}\else \v{S}\fi{}mejkal},
  \citenamefont {Marmodoro}, \citenamefont {Ahn}, \citenamefont
  {Gonz\'alez-Hern\'andez}, \citenamefont {Turek}, \citenamefont {Mankovsky},
  \citenamefont {Ebert}, \citenamefont {D'Souza}, \citenamefont
  {\ifmmode~\check{S}\else \v{S}\fi{}ipr}, \citenamefont {Sinova},\ and\
  \citenamefont {Jungwirth}}]{Libor2023}%
  \BibitemOpen
  \bibfield  {author} {\bibinfo {author} {\bibfnamefont {L.}~\bibnamefont
  {\ifmmode~\check{S}\else \v{S}\fi{}mejkal}}, \bibinfo {author} {\bibfnamefont
  {A.}~\bibnamefont {Marmodoro}}, \bibinfo {author} {\bibfnamefont {K.-H.}\
  \bibnamefont {Ahn}}, \bibinfo {author} {\bibfnamefont {R.}~\bibnamefont
  {Gonz\'alez-Hern\'andez}}, \bibinfo {author} {\bibfnamefont {I.}~\bibnamefont
  {Turek}}, \bibinfo {author} {\bibfnamefont {S.}~\bibnamefont {Mankovsky}},
  \bibinfo {author} {\bibfnamefont {H.}~\bibnamefont {Ebert}}, \bibinfo
  {author} {\bibfnamefont {S.~W.}\ \bibnamefont {D'Souza}}, \bibinfo {author}
  {\bibfnamefont {O.~c.~v.}\ \bibnamefont {\ifmmode~\check{S}\else
  \v{S}\fi{}ipr}}, \bibinfo {author} {\bibfnamefont {J.}~\bibnamefont
  {Sinova}},\ and\ \bibinfo {author} {\bibfnamefont {T.~c.~v.}\ \bibnamefont
  {Jungwirth}},\ }\bibfield  {title} {\bibinfo {title} {Chiral magnons in
  altermagnetic \ce{RuO2}},\ }\href
  {https://doi.org/10.1103/PhysRevLett.131.256703} {\bibfield  {journal}
  {\bibinfo  {journal} {Phys. Rev. Lett.}\ }\textbf {\bibinfo {volume} {131}},\
  \bibinfo {pages} {256703} (\bibinfo {year} {2023})}\BibitemShut {NoStop}%
\bibitem [{\citenamefont {McClarty}\ \emph {et~al.}(2024)\citenamefont
  {McClarty}, \citenamefont {Gukasov},\ and\ \citenamefont
  {Rau}}]{McClarty2024}%
  \BibitemOpen
  \bibfield  {author} {\bibinfo {author} {\bibfnamefont {P.~A.}\ \bibnamefont
  {McClarty}}, \bibinfo {author} {\bibfnamefont {A.}~\bibnamefont {Gukasov}},\
  and\ \bibinfo {author} {\bibfnamefont {J.~G.}\ \bibnamefont {Rau}},\ }\href
  {https://arxiv.org/abs/2410.10771} {\bibinfo {title} {Observing
  altermagnetism using polarized neutrons}} (\bibinfo {year} {2024}),\ \Eprint
  {https://arxiv.org/abs/2410.10771} {arXiv:2410.10771 [cond-mat.str-el]}
  \BibitemShut {NoStop}%
\bibitem [{\citenamefont {Liu}\ \emph {et~al.}(2024)\citenamefont {Liu},
  \citenamefont {Ozeki}, \citenamefont {Asai}, \citenamefont {Itoh},\ and\
  \citenamefont {Masuda}}]{Liu2024}%
  \BibitemOpen
  \bibfield  {author} {\bibinfo {author} {\bibfnamefont {Z.}~\bibnamefont
  {Liu}}, \bibinfo {author} {\bibfnamefont {M.}~\bibnamefont {Ozeki}}, \bibinfo
  {author} {\bibfnamefont {S.}~\bibnamefont {Asai}}, \bibinfo {author}
  {\bibfnamefont {S.}~\bibnamefont {Itoh}},\ and\ \bibinfo {author}
  {\bibfnamefont {T.}~\bibnamefont {Masuda}},\ }\bibfield  {title} {\bibinfo
  {title} {Chiral split magnon in altermagnetic {MnTe}},\ }\href
  {https://doi.org/10.1103/PhysRevLett.133.156702} {\bibfield  {journal}
  {\bibinfo  {journal} {Phys. Rev. Lett.}\ }\textbf {\bibinfo {volume} {133}},\
  \bibinfo {pages} {156702} (\bibinfo {year} {2024})}\BibitemShut {NoStop}%
\bibitem [{Note1()}]{Note1}%
  \BibitemOpen
  \bibinfo {note} {We will stick to handedness in the rest of the
  article}\BibitemShut {NoStop}%
\bibitem [{\citenamefont {Cui}\ \emph {et~al.}(2023)\citenamefont {Cui},
  \citenamefont {Zeng}, \citenamefont {Cui}, \citenamefont {Yu},\ and\
  \citenamefont {Yang}}]{Cui2023}%
  \BibitemOpen
  \bibfield  {author} {\bibinfo {author} {\bibfnamefont {Q.}~\bibnamefont
  {Cui}}, \bibinfo {author} {\bibfnamefont {B.}~\bibnamefont {Zeng}}, \bibinfo
  {author} {\bibfnamefont {P.}~\bibnamefont {Cui}}, \bibinfo {author}
  {\bibfnamefont {T.}~\bibnamefont {Yu}},\ and\ \bibinfo {author}
  {\bibfnamefont {H.}~\bibnamefont {Yang}},\ }\bibfield  {title} {\bibinfo
  {title} {Efficient spin seebeck and spin nernst effects of magnons in
  altermagnets},\ }\href {https://doi.org/10.1103/PhysRevB.108.L180401}
  {\bibfield  {journal} {\bibinfo  {journal} {Phys. Rev. B}\ }\textbf {\bibinfo
  {volume} {108}},\ \bibinfo {pages} {L180401} (\bibinfo {year}
  {2023})}\BibitemShut {NoStop}%
\bibitem [{\citenamefont {{de Haas}}\ \emph {et~al.}(1940)\citenamefont {{de
  Haas}}, \citenamefont {Schultz},\ and\ \citenamefont
  {Koolhaas}}]{deHaas1940}%
  \BibitemOpen
  \bibfield  {author} {\bibinfo {author} {\bibfnamefont {W.}~\bibnamefont {{de
  Haas}}}, \bibinfo {author} {\bibfnamefont {B.}~\bibnamefont {Schultz}},\ and\
  \bibinfo {author} {\bibfnamefont {J.}~\bibnamefont {Koolhaas}},\ }\bibfield
  {title} {\bibinfo {title} {Further measurements of the magnetic properties of
  some salts of the iron group at low temperatures},\ }\href
  {https://doi.org/https://doi.org/10.1016/S0031-8914(40)90069-6} {\bibfield
  {journal} {\bibinfo  {journal} {Physica}\ }\textbf {\bibinfo {volume} {7}},\
  \bibinfo {pages} {57} (\bibinfo {year} {1940})}\BibitemShut {NoStop}%
\bibitem [{\citenamefont {Seehra}\ and\ \citenamefont
  {Helmick}(1984)}]{Seehra1984}%
  \BibitemOpen
  \bibfield  {author} {\bibinfo {author} {\bibfnamefont {M.~S.}\ \bibnamefont
  {Seehra}}\ and\ \bibinfo {author} {\bibfnamefont {R.~E.}\ \bibnamefont
  {Helmick}},\ }\bibfield  {title} {\bibinfo {title} {Anomalous changes in the
  dielectric constants of {MnF$_2$} near its néel temperature},\ }\href
  {https://doi.org/10.1063/1.333652} {\bibfield  {journal} {\bibinfo  {journal}
  {J. Appl. Phys.}\ }\textbf {\bibinfo {volume} {55}},\ \bibinfo {pages} {2330}
  (\bibinfo {year} {1984})}\BibitemShut {NoStop}%
\bibitem [{\citenamefont {Yamani}\ \emph {et~al.}(2010)\citenamefont {Yamani},
  \citenamefont {Tun},\ and\ \citenamefont {Ryan}}]{Yamani2010}%
  \BibitemOpen
  \bibfield  {author} {\bibinfo {author} {\bibfnamefont {Z.}~\bibnamefont
  {Yamani}}, \bibinfo {author} {\bibfnamefont {Z.}~\bibnamefont {Tun}},\ and\
  \bibinfo {author} {\bibfnamefont {D.~H.}\ \bibnamefont {Ryan}},\ }\bibfield
  {title} {\bibinfo {title} {Neutron scattering study of the classical
  antiferromagnet {MnF$_2$}: a perfect hands-on neutron scattering teaching
  course},\ }\href {https://doi.org/10.1139/P10-081} {\bibfield  {journal}
  {\bibinfo  {journal} {Can. J. Phys.}\ }\textbf {\bibinfo {volume} {88}},\
  \bibinfo {pages} {771} (\bibinfo {year} {2010})}\BibitemShut {NoStop}%
\bibitem [{\citenamefont {Bhowal}\ and\ \citenamefont
  {Spaldin}(2024)}]{BhowalSpaldin2024}%
  \BibitemOpen
  \bibfield  {author} {\bibinfo {author} {\bibfnamefont {S.}~\bibnamefont
  {Bhowal}}\ and\ \bibinfo {author} {\bibfnamefont {N.~A.}\ \bibnamefont
  {Spaldin}},\ }\bibfield  {title} {\bibinfo {title} {Ferroically ordered
  magnetic octupoles in $d$-wave altermagnets},\ }\href
  {https://doi.org/10.1103/PhysRevX.14.011019} {\bibfield  {journal} {\bibinfo
  {journal} {Phys. Rev. X}\ }\textbf {\bibinfo {volume} {14}},\ \bibinfo
  {pages} {011019} (\bibinfo {year} {2024})}\BibitemShut {NoStop}%
\bibitem [{\citenamefont {Costa}\ and\ \citenamefont
  {Brown}(1989)}]{Costa1989}%
  \BibitemOpen
  \bibfield  {author} {\bibinfo {author} {\bibfnamefont {M.}~\bibnamefont
  {Costa}}\ and\ \bibinfo {author} {\bibfnamefont {P.}~\bibnamefont {Brown}},\
  }\bibfield  {title} {\bibinfo {title} {Magnetisation density in \ce{MnF2}},\
  }\href {https://doi.org/https://doi.org/10.1016/0921-4526(89)90669-8}
  {\bibfield  {journal} {\bibinfo  {journal} {Physica B: Condensed Matter}\
  }\textbf {\bibinfo {volume} {156-157}},\ \bibinfo {pages} {329} (\bibinfo
  {year} {1989})}\BibitemShut {NoStop}%
\bibitem [{\citenamefont {Mashkovich}\ \emph {et~al.}(2021)\citenamefont
  {Mashkovich}, \citenamefont {Grishunin}, \citenamefont {Dubrovin},
  \citenamefont {Zvezdin}, \citenamefont {Pisarev},\ and\ \citenamefont
  {Kimel}}]{Mashkovich2021}%
  \BibitemOpen
  \bibfield  {author} {\bibinfo {author} {\bibfnamefont {E.~A.}\ \bibnamefont
  {Mashkovich}}, \bibinfo {author} {\bibfnamefont {K.~A.}\ \bibnamefont
  {Grishunin}}, \bibinfo {author} {\bibfnamefont {R.~M.}\ \bibnamefont
  {Dubrovin}}, \bibinfo {author} {\bibfnamefont {A.~K.}\ \bibnamefont
  {Zvezdin}}, \bibinfo {author} {\bibfnamefont {R.~V.}\ \bibnamefont
  {Pisarev}},\ and\ \bibinfo {author} {\bibfnamefont {A.~V.}\ \bibnamefont
  {Kimel}},\ }\bibfield  {title} {\bibinfo {title} {Terahertz light–driven
  coupling of antiferromagnetic spins to lattice},\ }\href
  {https://doi.org/10.1126/science.abk1121} {\bibfield  {journal} {\bibinfo
  {journal} {Science}\ }\textbf {\bibinfo {volume} {374}},\ \bibinfo {pages}
  {1608} (\bibinfo {year} {2021})}\BibitemShut {NoStop}%
\bibitem [{\citenamefont {Disa}\ \emph {et~al.}(2020)\citenamefont {Disa},
  \citenamefont {Fechner}, \citenamefont {Nova}, \citenamefont {Liu},
  \citenamefont {F{\"o}rst}, \citenamefont {Prabhakaran}, \citenamefont
  {Radaelli},\ and\ \citenamefont {Cavalleri}}]{Disa2020}%
  \BibitemOpen
  \bibfield  {author} {\bibinfo {author} {\bibfnamefont {A.~S.}\ \bibnamefont
  {Disa}}, \bibinfo {author} {\bibfnamefont {M.}~\bibnamefont {Fechner}},
  \bibinfo {author} {\bibfnamefont {T.~F.}\ \bibnamefont {Nova}}, \bibinfo
  {author} {\bibfnamefont {B.}~\bibnamefont {Liu}}, \bibinfo {author}
  {\bibfnamefont {M.}~\bibnamefont {F{\"o}rst}}, \bibinfo {author}
  {\bibfnamefont {D.}~\bibnamefont {Prabhakaran}}, \bibinfo {author}
  {\bibfnamefont {P.~G.}\ \bibnamefont {Radaelli}},\ and\ \bibinfo {author}
  {\bibfnamefont {A.}~\bibnamefont {Cavalleri}},\ }\bibfield  {title} {\bibinfo
  {title} {Polarizing an antiferromagnet by optical engineering of the crystal
  field},\ }\href {https://doi.org/10.1038/s41567-020-0936-3} {\bibfield
  {journal} {\bibinfo  {journal} {Nat. Phys.}\ }\textbf {\bibinfo {volume}
  {16}},\ \bibinfo {pages} {937} (\bibinfo {year} {2020})}\BibitemShut
  {NoStop}%
\bibitem [{\citenamefont {Metzger}\ \emph {et~al.}(2024)\citenamefont
  {Metzger}, \citenamefont {Grishunin}, \citenamefont {Reinhoffer},
  \citenamefont {Dubrovin}, \citenamefont {Arshad}, \citenamefont {Ilyakov},
  \citenamefont {de~Oliveira}, \citenamefont {Ponomaryov}, \citenamefont
  {Deinert}, \citenamefont {Kovalev}, \citenamefont {Pisarev}, \citenamefont
  {Katsnelson}, \citenamefont {Ivanov}, \citenamefont {van Loosdrecht},
  \citenamefont {Kimel},\ and\ \citenamefont {Mashkovich}}]{Metzger2024}%
  \BibitemOpen
  \bibfield  {author} {\bibinfo {author} {\bibfnamefont {T.~W.~J.}\
  \bibnamefont {Metzger}}, \bibinfo {author} {\bibfnamefont {K.~A.}\
  \bibnamefont {Grishunin}}, \bibinfo {author} {\bibfnamefont {C.}~\bibnamefont
  {Reinhoffer}}, \bibinfo {author} {\bibfnamefont {R.~M.}\ \bibnamefont
  {Dubrovin}}, \bibinfo {author} {\bibfnamefont {A.}~\bibnamefont {Arshad}},
  \bibinfo {author} {\bibfnamefont {I.}~\bibnamefont {Ilyakov}}, \bibinfo
  {author} {\bibfnamefont {T.~V. A.~G.}\ \bibnamefont {de~Oliveira}}, \bibinfo
  {author} {\bibfnamefont {A.}~\bibnamefont {Ponomaryov}}, \bibinfo {author}
  {\bibfnamefont {J.-C.}\ \bibnamefont {Deinert}}, \bibinfo {author}
  {\bibfnamefont {S.}~\bibnamefont {Kovalev}}, \bibinfo {author} {\bibfnamefont
  {R.~V.}\ \bibnamefont {Pisarev}}, \bibinfo {author} {\bibfnamefont {M.~I.}\
  \bibnamefont {Katsnelson}}, \bibinfo {author} {\bibfnamefont {B.~A.}\
  \bibnamefont {Ivanov}}, \bibinfo {author} {\bibfnamefont {P.~H.~M.}\
  \bibnamefont {van Loosdrecht}}, \bibinfo {author} {\bibfnamefont {A.~V.}\
  \bibnamefont {Kimel}},\ and\ \bibinfo {author} {\bibfnamefont {E.~A.}\
  \bibnamefont {Mashkovich}},\ }\bibfield  {title} {\bibinfo {title}
  {Magnon-phonon fermi resonance in antiferromagnetic cof2},\ }\href
  {https://doi.org/10.1038/s41467-024-49716-w} {\bibfield  {journal} {\bibinfo
  {journal} {Nat. Commun.}\ }\textbf {\bibinfo {volume} {15}},\ \bibinfo
  {pages} {5472} (\bibinfo {year} {2024})}\BibitemShut {NoStop}%
\bibitem [{\citenamefont {Dubrovin}\ \emph {et~al.}(2024)\citenamefont
  {Dubrovin}, \citenamefont {Tellez-Mora}, \citenamefont {Garcia-Castro},
  \citenamefont {Siverin}, \citenamefont {Novikova}, \citenamefont {Boldyrev},
  \citenamefont {Mashkovich}, \citenamefont {Romero},\ and\ \citenamefont
  {Pisarev}}]{Dubrovin2024}%
  \BibitemOpen
  \bibfield  {author} {\bibinfo {author} {\bibfnamefont {R.~M.}\ \bibnamefont
  {Dubrovin}}, \bibinfo {author} {\bibfnamefont {A.}~\bibnamefont
  {Tellez-Mora}}, \bibinfo {author} {\bibfnamefont {A.~C.}\ \bibnamefont
  {Garcia-Castro}}, \bibinfo {author} {\bibfnamefont {N.~V.}\ \bibnamefont
  {Siverin}}, \bibinfo {author} {\bibfnamefont {N.~N.}\ \bibnamefont
  {Novikova}}, \bibinfo {author} {\bibfnamefont {K.~N.}\ \bibnamefont
  {Boldyrev}}, \bibinfo {author} {\bibfnamefont {E.~A.}\ \bibnamefont
  {Mashkovich}}, \bibinfo {author} {\bibfnamefont {A.~H.}\ \bibnamefont
  {Romero}},\ and\ \bibinfo {author} {\bibfnamefont {R.~V.}\ \bibnamefont
  {Pisarev}},\ }\bibfield  {title} {\bibinfo {title} {Polar phonons and
  magnetic excitations in the antiferromagnet \ce{CoF2}},\ }\href
  {https://doi.org/10.1103/PhysRevB.109.224312} {\bibfield  {journal} {\bibinfo
   {journal} {Phys. Rev. B}\ }\textbf {\bibinfo {volume} {109}},\ \bibinfo
  {pages} {224312} (\bibinfo {year} {2024})}\BibitemShut {NoStop}%
\bibitem [{\citenamefont {Duan}\ \emph {et~al.}(2025)\citenamefont {Duan},
  \citenamefont {Zhang}, \citenamefont {Zhu}, \citenamefont {Liu},
  \citenamefont {Zhang}, \citenamefont {\ifmmode \check{Z}\else
  \v{Z}\fi{}uti\ifmmode~\acute{c}\else \'{c}\fi{}},\ and\ \citenamefont
  {Zhou}}]{Duan2025}%
  \BibitemOpen
  \bibfield  {author} {\bibinfo {author} {\bibfnamefont {X.}~\bibnamefont
  {Duan}}, \bibinfo {author} {\bibfnamefont {J.}~\bibnamefont {Zhang}},
  \bibinfo {author} {\bibfnamefont {Z.}~\bibnamefont {Zhu}}, \bibinfo {author}
  {\bibfnamefont {Y.}~\bibnamefont {Liu}}, \bibinfo {author} {\bibfnamefont
  {Z.}~\bibnamefont {Zhang}}, \bibinfo {author} {\bibfnamefont
  {I.}~\bibnamefont {\ifmmode \check{Z}\else
  \v{Z}\fi{}uti\ifmmode~\acute{c}\else \'{c}\fi{}}},\ and\ \bibinfo {author}
  {\bibfnamefont {T.}~\bibnamefont {Zhou}},\ }\bibfield  {title} {\bibinfo
  {title} {Antiferroelectric altermagnets: Antiferroelectricity alters
  magnets},\ }\href {https://doi.org/10.1103/PhysRevLett.134.106801} {\bibfield
   {journal} {\bibinfo  {journal} {Phys. Rev. Lett.}\ }\textbf {\bibinfo
  {volume} {134}},\ \bibinfo {pages} {106801} (\bibinfo {year}
  {2025})}\BibitemShut {NoStop}%
\bibitem [{\citenamefont {Gu}\ \emph {et~al.}(2025)\citenamefont {Gu},
  \citenamefont {Liu}, \citenamefont {Zhu}, \citenamefont {Yananose},
  \citenamefont {Chen}, \citenamefont {Hu}, \citenamefont {Stroppa},\ and\
  \citenamefont {Liu}}]{Gu2025}%
  \BibitemOpen
  \bibfield  {author} {\bibinfo {author} {\bibfnamefont {M.}~\bibnamefont
  {Gu}}, \bibinfo {author} {\bibfnamefont {Y.}~\bibnamefont {Liu}}, \bibinfo
  {author} {\bibfnamefont {H.}~\bibnamefont {Zhu}}, \bibinfo {author}
  {\bibfnamefont {K.}~\bibnamefont {Yananose}}, \bibinfo {author}
  {\bibfnamefont {X.}~\bibnamefont {Chen}}, \bibinfo {author} {\bibfnamefont
  {Y.}~\bibnamefont {Hu}}, \bibinfo {author} {\bibfnamefont {A.}~\bibnamefont
  {Stroppa}},\ and\ \bibinfo {author} {\bibfnamefont {Q.}~\bibnamefont {Liu}},\
  }\bibfield  {title} {\bibinfo {title} {Ferroelectric switchable
  altermagnetism},\ }\href {https://doi.org/10.1103/PhysRevLett.134.106802}
  {\bibfield  {journal} {\bibinfo  {journal} {Phys. Rev. Lett.}\ }\textbf
  {\bibinfo {volume} {134}},\ \bibinfo {pages} {106802} (\bibinfo {year}
  {2025})}\BibitemShut {NoStop}%
\bibitem [{\citenamefont {Šmejkal}(2024)}]{Libor2025}%
  \BibitemOpen
  \bibfield  {author} {\bibinfo {author} {\bibfnamefont {L.}~\bibnamefont
  {Šmejkal}},\ }\href {https://arxiv.org/abs/2411.19928} {\bibinfo {title}
  {Altermagnetic multiferroics and altermagnetoelectric effect}} (\bibinfo
  {year} {2024}),\ \Eprint {https://arxiv.org/abs/2411.19928} {arXiv:2411.19928
  [cond-mat.mtrl-sci]} \BibitemShut {NoStop}%
\bibitem [{\citenamefont {Kirilyuk}\ \emph {et~al.}(2010)\citenamefont
  {Kirilyuk}, \citenamefont {Kimel},\ and\ \citenamefont
  {Rasing}}]{Kirilyuk2010}%
  \BibitemOpen
  \bibfield  {author} {\bibinfo {author} {\bibfnamefont {A.}~\bibnamefont
  {Kirilyuk}}, \bibinfo {author} {\bibfnamefont {A.~V.}\ \bibnamefont
  {Kimel}},\ and\ \bibinfo {author} {\bibfnamefont {T.}~\bibnamefont
  {Rasing}},\ }\bibfield  {title} {\bibinfo {title} {Ultrafast optical
  manipulation of magnetic order},\ }\href
  {https://doi.org/10.1103/RevModPhys.82.2731} {\bibfield  {journal} {\bibinfo
  {journal} {Rev. Mod. Phys.}\ }\textbf {\bibinfo {volume} {82}},\ \bibinfo
  {pages} {2731} (\bibinfo {year} {2010})}\BibitemShut {NoStop}%
\bibitem [{\citenamefont {Bl\"ochl}(1994)}]{Bloch1994}%
  \BibitemOpen
  \bibfield  {author} {\bibinfo {author} {\bibfnamefont {P.~E.}\ \bibnamefont
  {Bl\"ochl}},\ }\bibfield  {title} {\bibinfo {title} {Projector augmented-wave
  method},\ }\href {https://doi.org/10.1103/PhysRevB.50.17953} {\bibfield
  {journal} {\bibinfo  {journal} {Phys. Rev. B}\ }\textbf {\bibinfo {volume}
  {50}},\ \bibinfo {pages} {17953} (\bibinfo {year} {1994})}\BibitemShut
  {NoStop}%
\bibitem [{\citenamefont {Kresse}\ and\ \citenamefont
  {Joubert}(1999)}]{Kresse1999}%
  \BibitemOpen
  \bibfield  {author} {\bibinfo {author} {\bibfnamefont {G.}~\bibnamefont
  {Kresse}}\ and\ \bibinfo {author} {\bibfnamefont {D.}~\bibnamefont
  {Joubert}},\ }\bibfield  {title} {\bibinfo {title} {From ultrasoft
  pseudopotentials to the projector augmented-wave method},\ }\href
  {https://doi.org/10.1103/PhysRevB.59.1758} {\bibfield  {journal} {\bibinfo
  {journal} {Phys. Rev. B}\ }\textbf {\bibinfo {volume} {59}},\ \bibinfo
  {pages} {1758} (\bibinfo {year} {1999})}\BibitemShut {NoStop}%
\bibitem [{\citenamefont {Kresse}\ and\ \citenamefont
  {Hafner}(1993)}]{Kresse1993}%
  \BibitemOpen
  \bibfield  {author} {\bibinfo {author} {\bibfnamefont {G.}~\bibnamefont
  {Kresse}}\ and\ \bibinfo {author} {\bibfnamefont {J.}~\bibnamefont
  {Hafner}},\ }\bibfield  {title} {\bibinfo {title} {Ab initio molecular
  dynamics for liquid metals},\ }\href
  {https://doi.org/10.1103/PhysRevB.47.558} {\bibfield  {journal} {\bibinfo
  {journal} {Phys. Rev. B}\ }\textbf {\bibinfo {volume} {47}},\ \bibinfo
  {pages} {558} (\bibinfo {year} {1993})}\BibitemShut {NoStop}%
\bibitem [{\citenamefont {Kresse}\ and\ \citenamefont
  {Furthm\"uller}(1996)}]{Kresse1996}%
  \BibitemOpen
  \bibfield  {author} {\bibinfo {author} {\bibfnamefont {G.}~\bibnamefont
  {Kresse}}\ and\ \bibinfo {author} {\bibfnamefont {J.}~\bibnamefont
  {Furthm\"uller}},\ }\bibfield  {title} {\bibinfo {title} {Efficient iterative
  schemes for ab initio total-energy calculations using a plane-wave basis
  set},\ }\href {https://doi.org/10.1103/PhysRevB.54.11169} {\bibfield
  {journal} {\bibinfo  {journal} {Phys. Rev. B}\ }\textbf {\bibinfo {volume}
  {54}},\ \bibinfo {pages} {11169} (\bibinfo {year} {1996})}\BibitemShut
  {NoStop}%
\bibitem [{\citenamefont {Xiang}\ \emph {et~al.}(2011)\citenamefont {Xiang},
  \citenamefont {Kan}, \citenamefont {Wei}, \citenamefont {Whangbo},\ and\
  \citenamefont {Gong}}]{Xiang2011}%
  \BibitemOpen
  \bibfield  {author} {\bibinfo {author} {\bibfnamefont {H.~J.}\ \bibnamefont
  {Xiang}}, \bibinfo {author} {\bibfnamefont {E.~J.}\ \bibnamefont {Kan}},
  \bibinfo {author} {\bibfnamefont {S.-H.}\ \bibnamefont {Wei}}, \bibinfo
  {author} {\bibfnamefont {M.-H.}\ \bibnamefont {Whangbo}},\ and\ \bibinfo
  {author} {\bibfnamefont {X.~G.}\ \bibnamefont {Gong}},\ }\bibfield  {title}
  {\bibinfo {title} {Predicting the spin-lattice order of frustrated systems
  from first principles},\ }\href {https://doi.org/10.1103/PhysRevB.84.224429}
  {\bibfield  {journal} {\bibinfo  {journal} {Phys. Rev. B}\ }\textbf {\bibinfo
  {volume} {84}},\ \bibinfo {pages} {224429} (\bibinfo {year}
  {2011})}\BibitemShut {NoStop}%
\bibitem [{\citenamefont {\ifmmode~\check{S}\else \v{S}\fi{}abani}\ \emph
  {et~al.}(2020)\citenamefont {\ifmmode~\check{S}\else \v{S}\fi{}abani},
  \citenamefont {Bacaksiz},\ and\ \citenamefont {Milo\ifmmode \check{s}\else
  \v{s}\fi{}evi\ifmmode~\acute{c}\else \'{c}\fi{}}}]{Sabani2020}%
  \BibitemOpen
  \bibfield  {author} {\bibinfo {author} {\bibfnamefont {D.}~\bibnamefont
  {\ifmmode~\check{S}\else \v{S}\fi{}abani}}, \bibinfo {author} {\bibfnamefont
  {C.}~\bibnamefont {Bacaksiz}},\ and\ \bibinfo {author} {\bibfnamefont
  {M.~V.}\ \bibnamefont {Milo\ifmmode \check{s}\else
  \v{s}\fi{}evi\ifmmode~\acute{c}\else \'{c}\fi{}}},\ }\bibfield  {title}
  {\bibinfo {title} {Ab initio methodology for magnetic exchange parameters:
  Generic four-state energy mapping onto a heisenberg spin hamiltonian},\
  }\href {https://doi.org/10.1103/PhysRevB.102.014457} {\bibfield  {journal}
  {\bibinfo  {journal} {Phys. Rev. B}\ }\textbf {\bibinfo {volume} {102}},\
  \bibinfo {pages} {014457} (\bibinfo {year} {2020})}\BibitemShut {NoStop}%
\bibitem [{SM()}]{SM}%
  \BibitemOpen
  \href@noop {} {\bibinfo {title} {Supplemental materials for the computational
  details of the {DFT} and the magnon calculations, and the additional results
  for the exchange coupling calculations.}}\BibitemShut {Stop}%
\bibitem [{not()}]{note}%
  \BibitemOpen
  \href@noop {} {}\bibinfo {note} {{\color{black}Although Ref. \cite{Libor2023}
  considers an antiferromagnetic ground state, the ground state magnetism of
  {RuO$_2$} remains debatable \cite{Smolyanyuk2024,
  Hiraishi2024,Philipp2024}}.}\BibitemShut {Stop}%
\bibitem [{\citenamefont {Smolyanyuk}\ \emph {et~al.}(2024)\citenamefont
  {Smolyanyuk}, \citenamefont {Mazin}, \citenamefont {Garcia-Gassull},\ and\
  \citenamefont {Valent\'{\i}}}]{Smolyanyuk2024}%
  \BibitemOpen
  \bibfield  {author} {\bibinfo {author} {\bibfnamefont {A.}~\bibnamefont
  {Smolyanyuk}}, \bibinfo {author} {\bibfnamefont {I.~I.}\ \bibnamefont
  {Mazin}}, \bibinfo {author} {\bibfnamefont {L.}~\bibnamefont
  {Garcia-Gassull}},\ and\ \bibinfo {author} {\bibfnamefont {R.}~\bibnamefont
  {Valent\'{\i}}},\ }\bibfield  {title} {\bibinfo {title} {Fragility of the
  magnetic order in the prototypical altermagnet {RuO$_2$}},\ }\href
  {https://doi.org/10.1103/PhysRevB.109.134424} {\bibfield  {journal} {\bibinfo
   {journal} {Phys. Rev. B}\ }\textbf {\bibinfo {volume} {109}},\ \bibinfo
  {pages} {134424} (\bibinfo {year} {2024})}\BibitemShut {NoStop}%
\bibitem [{\citenamefont {Hiraishi}\ \emph {et~al.}(2024)\citenamefont
  {Hiraishi}, \citenamefont {Okabe}, \citenamefont {Koda}, \citenamefont
  {Kadono}, \citenamefont {Muroi}, \citenamefont {Hirai},\ and\ \citenamefont
  {Hiroi}}]{Hiraishi2024}%
  \BibitemOpen
  \bibfield  {author} {\bibinfo {author} {\bibfnamefont {M.}~\bibnamefont
  {Hiraishi}}, \bibinfo {author} {\bibfnamefont {H.}~\bibnamefont {Okabe}},
  \bibinfo {author} {\bibfnamefont {A.}~\bibnamefont {Koda}}, \bibinfo {author}
  {\bibfnamefont {R.}~\bibnamefont {Kadono}}, \bibinfo {author} {\bibfnamefont
  {T.}~\bibnamefont {Muroi}}, \bibinfo {author} {\bibfnamefont
  {D.}~\bibnamefont {Hirai}},\ and\ \bibinfo {author} {\bibfnamefont
  {Z.}~\bibnamefont {Hiroi}},\ }\bibfield  {title} {\bibinfo {title}
  {Nonmagnetic ground state in {RuO$_2$} revealed by muon spin rotation},\
  }\href {https://doi.org/10.1103/PhysRevLett.132.166702} {\bibfield  {journal}
  {\bibinfo  {journal} {Phys. Rev. Lett.}\ }\textbf {\bibinfo {volume} {132}},\
  \bibinfo {pages} {166702} (\bibinfo {year} {2024})}\BibitemShut {NoStop}%
\bibitem [{\citenamefont {Ke{\ss}ler}\ \emph {et~al.}(2024)\citenamefont
  {Ke{\ss}ler}, \citenamefont {Garcia-Gassull}, \citenamefont {Suter},
  \citenamefont {Prokscha}, \citenamefont {Salman}, \citenamefont {Khalyavin},
  \citenamefont {Manuel}, \citenamefont {Orlandi}, \citenamefont {Mazin},
  \citenamefont {Valent{\'\i}},\ and\ \citenamefont {Moser}}]{Philipp2024}%
  \BibitemOpen
  \bibfield  {author} {\bibinfo {author} {\bibfnamefont {P.}~\bibnamefont
  {Ke{\ss}ler}}, \bibinfo {author} {\bibfnamefont {L.}~\bibnamefont
  {Garcia-Gassull}}, \bibinfo {author} {\bibfnamefont {A.}~\bibnamefont
  {Suter}}, \bibinfo {author} {\bibfnamefont {T.}~\bibnamefont {Prokscha}},
  \bibinfo {author} {\bibfnamefont {Z.}~\bibnamefont {Salman}}, \bibinfo
  {author} {\bibfnamefont {D.}~\bibnamefont {Khalyavin}}, \bibinfo {author}
  {\bibfnamefont {P.}~\bibnamefont {Manuel}}, \bibinfo {author} {\bibfnamefont
  {F.}~\bibnamefont {Orlandi}}, \bibinfo {author} {\bibfnamefont {I.~I.}\
  \bibnamefont {Mazin}}, \bibinfo {author} {\bibfnamefont {R.}~\bibnamefont
  {Valent{\'\i}}},\ and\ \bibinfo {author} {\bibfnamefont {S.}~\bibnamefont
  {Moser}},\ }\bibfield  {title} {\bibinfo {title} {Absence of magnetic order
  in {RuO$_2$}: insights from $\mu$sr spectroscopy and neutron diffraction},\
  }\href {https://doi.org/10.1038/s44306-024-00055-y} {\bibfield  {journal}
  {\bibinfo  {journal} {npj Spintronics}\ }\textbf {\bibinfo {volume} {2}},\
  \bibinfo {pages} {50} (\bibinfo {year} {2024})}\BibitemShut {NoStop}%
\bibitem [{\citenamefont {Holstein}\ and\ \citenamefont
  {Primakoff}(1940)}]{HolsteinPrimakoff1940}%
  \BibitemOpen
  \bibfield  {author} {\bibinfo {author} {\bibfnamefont {T.}~\bibnamefont
  {Holstein}}\ and\ \bibinfo {author} {\bibfnamefont {H.}~\bibnamefont
  {Primakoff}},\ }\bibfield  {title} {\bibinfo {title} {Field dependence of the
  intrinsic domain magnetization of a ferromagnet},\ }\href
  {https://doi.org/10.1103/PhysRev.58.1098} {\bibfield  {journal} {\bibinfo
  {journal} {Phys. Rev.}\ }\textbf {\bibinfo {volume} {58}},\ \bibinfo {pages}
  {1098} (\bibinfo {year} {1940})}\BibitemShut {NoStop}%
\bibitem [{\citenamefont {Morano}\ \emph {et~al.}(2024)\citenamefont {Morano},
  \citenamefont {Maesen}, \citenamefont {Nikitin}, \citenamefont {Lass},
  \citenamefont {Mazzone},\ and\ \citenamefont {Zaharko}}]{Morano2024}%
  \BibitemOpen
  \bibfield  {author} {\bibinfo {author} {\bibfnamefont {V.~C.}\ \bibnamefont
  {Morano}}, \bibinfo {author} {\bibfnamefont {Z.}~\bibnamefont {Maesen}},
  \bibinfo {author} {\bibfnamefont {S.~E.}\ \bibnamefont {Nikitin}}, \bibinfo
  {author} {\bibfnamefont {J.}~\bibnamefont {Lass}}, \bibinfo {author}
  {\bibfnamefont {D.~G.}\ \bibnamefont {Mazzone}},\ and\ \bibinfo {author}
  {\bibfnamefont {O.}~\bibnamefont {Zaharko}},\ }\href
  {https://arxiv.org/abs/2412.03545} {\bibinfo {title} {Absence of
  altermagnetic magnon band splitting in \ce{MnF2}}} (\bibinfo {year} {2024}),\
  \Eprint {https://arxiv.org/abs/2412.03545} {arXiv:2412.03545
  [cond-mat.str-el]} \BibitemShut {NoStop}%
\bibitem [{\citenamefont {Togo}\ and\ \citenamefont {Tanaka}(2015)}]{Phonopy}%
  \BibitemOpen
  \bibfield  {author} {\bibinfo {author} {\bibfnamefont {A.}~\bibnamefont
  {Togo}}\ and\ \bibinfo {author} {\bibfnamefont {I.}~\bibnamefont {Tanaka}},\
  }\bibfield  {title} {\bibinfo {title} {First principles phonon calculations
  in materials science},\ }\href
  {https://doi.org/https://doi.org/10.1016/j.scriptamat.2015.07.021} {\bibfield
   {journal} {\bibinfo  {journal} {Scripta Materialia}\ }\textbf {\bibinfo
  {volume} {108}},\ \bibinfo {pages} {1} (\bibinfo {year} {2015})}\BibitemShut
  {NoStop}%
\bibitem [{\citenamefont {Stokes}\ \emph {et~al.}({\natexlab{a}})\citenamefont
  {Stokes}, \citenamefont {Hatch},\ and\ \citenamefont
  {Campbell}}]{Isodistort}%
  \BibitemOpen
  \bibfield  {author} {\bibinfo {author} {\bibfnamefont {H.~T.}\ \bibnamefont
  {Stokes}}, \bibinfo {author} {\bibfnamefont {D.~M.}\ \bibnamefont {Hatch}},\
  and\ \bibinfo {author} {\bibfnamefont {B.~J.}\ \bibnamefont {Campbell}},\
  }\href@noop {} {\bibinfo {title} {{ISODISTORT, ISOTROPY} software suite}},\
  \bibinfo {howpublished} {\url{https://iso.byu.edu}}
  ({\natexlab{a}})\BibitemShut {NoStop}%
\bibitem [{\citenamefont {Campbell}\ \emph {et~al.}(2006)\citenamefont
  {Campbell}, \citenamefont {Stokes}, \citenamefont {Tanner},\ and\
  \citenamefont {Hatch}}]{Isodistort1}%
  \BibitemOpen
  \bibfield  {author} {\bibinfo {author} {\bibfnamefont {B.~J.}\ \bibnamefont
  {Campbell}}, \bibinfo {author} {\bibfnamefont {H.~T.}\ \bibnamefont
  {Stokes}}, \bibinfo {author} {\bibfnamefont {D.~E.}\ \bibnamefont {Tanner}},\
  and\ \bibinfo {author} {\bibfnamefont {D.~M.}\ \bibnamefont {Hatch}},\
  }\bibfield  {title} {\bibinfo {title} {{{\it ISODISPLACE}: a web-based tool
  for exploring structural distortions}},\ }\href@noop {} {\bibfield  {journal}
  {\bibinfo  {journal} {Journal of Applied Crystallography}\ }\textbf {\bibinfo
  {volume} {39}},\ \bibinfo {pages} {607} (\bibinfo {year} {2006})}\BibitemShut
  {NoStop}%
\bibitem [{\citenamefont {Bandyopadhyay}\ and\ \citenamefont
  {Ghosez}(2024)}]{SB_TMO}%
  \BibitemOpen
  \bibfield  {author} {\bibinfo {author} {\bibfnamefont {S.}~\bibnamefont
  {Bandyopadhyay}}\ and\ \bibinfo {author} {\bibfnamefont {P.}~\bibnamefont
  {Ghosez}},\ }\href {https://arxiv.org/abs/2407.21406} {\bibinfo {title}
  {Structurally triggered orbital and charge orderings in \ce{TlMnO3} and
  related compounds}} (\bibinfo {year} {2024}),\ \Eprint
  {https://arxiv.org/abs/2407.21406} {arXiv:2407.21406 [cond-mat.str-el]}
  \BibitemShut {NoStop}%
\bibitem [{Note2()}]{Note2}%
  \BibitemOpen
  \bibinfo {note} {$Q_{A_{2u}}$ and $Q_{A_{1g}}$ are normalized to their
  amplitudes at $P4_2/nnm$ structure}\BibitemShut {NoStop}%
\bibitem [{\citenamefont {Hioki}\ \emph {et~al.}(2022)\citenamefont {Hioki},
  \citenamefont {Hashimoto},\ and\ \citenamefont {Saitoh}}]{Hioki2022}%
  \BibitemOpen
  \bibfield  {author} {\bibinfo {author} {\bibfnamefont {T.}~\bibnamefont
  {Hioki}}, \bibinfo {author} {\bibfnamefont {Y.}~\bibnamefont {Hashimoto}},\
  and\ \bibinfo {author} {\bibfnamefont {E.}~\bibnamefont {Saitoh}},\
  }\bibfield  {title} {\bibinfo {title} {Coherent oscillation between phonons
  and magnons},\ }\href {https://doi.org/10.1038/s42005-022-00888-1} {\bibfield
   {journal} {\bibinfo  {journal} {Communications Physics}\ }\textbf {\bibinfo
  {volume} {5}},\ \bibinfo {pages} {115} (\bibinfo {year} {2022})}\BibitemShut
  {NoStop}%
\bibitem [{\citenamefont {Liu}\ \emph {et~al.}(2021)\citenamefont {Liu},
  \citenamefont {Granados~del \'Aguila}, \citenamefont {Bhowmick},
  \citenamefont {Gan}, \citenamefont {Thu Ha~Do}, \citenamefont {Prosnikov},
  \citenamefont {Sedmidubsk\'y}, \citenamefont {Sofer}, \citenamefont
  {Christianen}, \citenamefont {Sengupta},\ and\ \citenamefont
  {Xiong}}]{Liu2021}%
  \BibitemOpen
  \bibfield  {author} {\bibinfo {author} {\bibfnamefont {S.}~\bibnamefont
  {Liu}}, \bibinfo {author} {\bibfnamefont {A.}~\bibnamefont {Granados~del
  \'Aguila}}, \bibinfo {author} {\bibfnamefont {D.}~\bibnamefont {Bhowmick}},
  \bibinfo {author} {\bibfnamefont {C.~K.}\ \bibnamefont {Gan}}, \bibinfo
  {author} {\bibfnamefont {T.}~\bibnamefont {Thu Ha~Do}}, \bibinfo {author}
  {\bibfnamefont {M.~A.}\ \bibnamefont {Prosnikov}}, \bibinfo {author}
  {\bibfnamefont {D.}~\bibnamefont {Sedmidubsk\'y}}, \bibinfo {author}
  {\bibfnamefont {Z.}~\bibnamefont {Sofer}}, \bibinfo {author} {\bibfnamefont
  {P.~C.~M.}\ \bibnamefont {Christianen}}, \bibinfo {author} {\bibfnamefont
  {P.}~\bibnamefont {Sengupta}},\ and\ \bibinfo {author} {\bibfnamefont
  {Q.}~\bibnamefont {Xiong}},\ }\bibfield  {title} {\bibinfo {title} {Direct
  observation of magnon-phonon strong coupling in two-dimensional
  antiferromagnet at high magnetic fields},\ }\href
  {https://doi.org/10.1103/PhysRevLett.127.097401} {\bibfield  {journal}
  {\bibinfo  {journal} {Phys. Rev. Lett.}\ }\textbf {\bibinfo {volume} {127}},\
  \bibinfo {pages} {097401} (\bibinfo {year} {2021})}\BibitemShut {NoStop}%
\bibitem [{\citenamefont {Mai}\ \emph {et~al.}(2021)\citenamefont {Mai},
  \citenamefont {Garrity}, \citenamefont {McCreary}, \citenamefont {Argo},
  \citenamefont {Simpson}, \citenamefont {Doan-Nguyen}, \citenamefont
  {Aguilar},\ and\ \citenamefont {Walker}}]{Mai2021}%
  \BibitemOpen
  \bibfield  {author} {\bibinfo {author} {\bibfnamefont {T.~T.}\ \bibnamefont
  {Mai}}, \bibinfo {author} {\bibfnamefont {K.~F.}\ \bibnamefont {Garrity}},
  \bibinfo {author} {\bibfnamefont {A.}~\bibnamefont {McCreary}}, \bibinfo
  {author} {\bibfnamefont {J.}~\bibnamefont {Argo}}, \bibinfo {author}
  {\bibfnamefont {J.~R.}\ \bibnamefont {Simpson}}, \bibinfo {author}
  {\bibfnamefont {V.}~\bibnamefont {Doan-Nguyen}}, \bibinfo {author}
  {\bibfnamefont {R.~V.}\ \bibnamefont {Aguilar}},\ and\ \bibinfo {author}
  {\bibfnamefont {A.~R.~H.}\ \bibnamefont {Walker}},\ }\bibfield  {title}
  {\bibinfo {title} {Magnon-phonon hybridization in 2d antiferromagnet
  {MnPSe$_3$}},\ }\href {https://doi.org/10.1126/sciadv.abj3106} {\bibfield
  {journal} {\bibinfo  {journal} {Science Advances}\ }\textbf {\bibinfo
  {volume} {7}},\ \bibinfo {pages} {eabj3106} (\bibinfo {year}
  {2021})}\BibitemShut {NoStop}%
\bibitem [{sym()}]{symm}%
  \BibitemOpen
  \href@noop {} {}\bibinfo {note} {The symmetry allowed terms in the free
  energy are obtained using the INVARIANTS software
  \cite{Invariants,Invariants2}}\BibitemShut {NoStop}%
\bibitem [{\citenamefont {Stokes}\ \emph {et~al.}({\natexlab{b}})\citenamefont
  {Stokes}, \citenamefont {Hatch},\ and\ \citenamefont
  {Campbell}}]{Invariants}%
  \BibitemOpen
  \bibfield  {author} {\bibinfo {author} {\bibfnamefont {H.~T.}\ \bibnamefont
  {Stokes}}, \bibinfo {author} {\bibfnamefont {D.~M.}\ \bibnamefont {Hatch}},\
  and\ \bibinfo {author} {\bibfnamefont {B.~J.}\ \bibnamefont {Campbell}},\
  }\href {https://iso.byu.edu} {\bibfield  {journal} {\bibinfo  {journal}
  {INVARIANTS, ISOTROPY Software Suite}\ } ({\natexlab{b}})}\BibitemShut
  {NoStop}%
\bibitem [{\citenamefont {Hatch}\ and\ \citenamefont
  {Stokes}(2003)}]{Invariants2}%
  \BibitemOpen
  \bibfield  {author} {\bibinfo {author} {\bibfnamefont {D.~M.}\ \bibnamefont
  {Hatch}}\ and\ \bibinfo {author} {\bibfnamefont {H.~T.}\ \bibnamefont
  {Stokes}},\ }\bibfield  {title} {\bibinfo {title} {{{\it INVARIANTS}: program
  for obtaining a list of invariant polynomials of the order-parameter
  components associated with irreducible representations of a space group}},\
  }\href {https://doi.org/10.1107/S0021889803005946} {\bibfield  {journal}
  {\bibinfo  {journal} {Journal of Applied Crystallography}\ }\textbf {\bibinfo
  {volume} {36}},\ \bibinfo {pages} {951} (\bibinfo {year} {2003})}\BibitemShut
  {NoStop}%
\bibitem [{\citenamefont {Kim}\ \emph {et~al.}(2023)\citenamefont {Kim},
  \citenamefont {Lee}, \citenamefont {Lee}, \citenamefont {Lee}, \citenamefont
  {Ryoo}, \citenamefont {Byun}, \citenamefont {Lee}, \citenamefont {Kim},
  \citenamefont {Park},\ and\ \citenamefont {Hwang}}]{Kyung_2023}%
  \BibitemOpen
  \bibfield  {author} {\bibinfo {author} {\bibfnamefont {K.~D.}\ \bibnamefont
  {Kim}}, \bibinfo {author} {\bibfnamefont {Y.~B.}\ \bibnamefont {Lee}},
  \bibinfo {author} {\bibfnamefont {S.~H.}\ \bibnamefont {Lee}}, \bibinfo
  {author} {\bibfnamefont {I.~S.}\ \bibnamefont {Lee}}, \bibinfo {author}
  {\bibfnamefont {S.~K.}\ \bibnamefont {Ryoo}}, \bibinfo {author}
  {\bibfnamefont {S.}~\bibnamefont {Byun}}, \bibinfo {author} {\bibfnamefont
  {J.~H.}\ \bibnamefont {Lee}}, \bibinfo {author} {\bibfnamefont
  {H.}~\bibnamefont {Kim}}, \bibinfo {author} {\bibfnamefont {H.~W.}\
  \bibnamefont {Park}},\ and\ \bibinfo {author} {\bibfnamefont {C.~S.}\
  \bibnamefont {Hwang}},\ }\bibfield  {title} {\bibinfo {title} {Evolution of
  the ferroelectric properties of alscn film by electrical cycling with an
  inhomogeneous field distribution},\ }\href
  {https://doi.org/https://doi.org/10.1002/aelm.202201142} {\bibfield
  {journal} {\bibinfo  {journal} {Advanced Electronic Materials}\ }\textbf
  {\bibinfo {volume} {9}},\ \bibinfo {pages} {2201142} (\bibinfo {year}
  {2023})}\BibitemShut {NoStop}%
\bibitem [{\citenamefont {Varignon}\ \emph {et~al.}(2016)\citenamefont
  {Varignon}, \citenamefont {Bristowe},\ and\ \citenamefont
  {Ghosez}}]{Varignon_2016}%
  \BibitemOpen
  \bibfield  {author} {\bibinfo {author} {\bibfnamefont {J.}~\bibnamefont
  {Varignon}}, \bibinfo {author} {\bibfnamefont {N.~C.}\ \bibnamefont
  {Bristowe}},\ and\ \bibinfo {author} {\bibfnamefont {P.}~\bibnamefont
  {Ghosez}},\ }\bibfield  {title} {\bibinfo {title} {Electric field control of
  jahn-teller distortions in bulk perovskites},\ }\href
  {https://doi.org/10.1103/PhysRevLett.116.057602} {\bibfield  {journal}
  {\bibinfo  {journal} {Phys. Rev. Lett.}\ }\textbf {\bibinfo {volume} {116}},\
  \bibinfo {pages} {057602} (\bibinfo {year} {2016})}\BibitemShut {NoStop}%
\bibitem [{\citenamefont {Yang}\ \emph {et~al.}(2014)\citenamefont {Yang},
  \citenamefont {\'I\~niguez}, \citenamefont {Mao},\ and\ \citenamefont
  {Bellaiche}}]{Bellaiche_2014}%
  \BibitemOpen
  \bibfield  {author} {\bibinfo {author} {\bibfnamefont {Y.}~\bibnamefont
  {Yang}}, \bibinfo {author} {\bibfnamefont {J.}~\bibnamefont {\'I\~niguez}},
  \bibinfo {author} {\bibfnamefont {A.-J.}\ \bibnamefont {Mao}},\ and\ \bibinfo
  {author} {\bibfnamefont {L.}~\bibnamefont {Bellaiche}},\ }\bibfield  {title}
  {\bibinfo {title} {Prediction of a novel magnetoelectric switching mechanism
  in multiferroics},\ }\href {https://doi.org/10.1103/PhysRevLett.112.057202}
  {\bibfield  {journal} {\bibinfo  {journal} {Phys. Rev. Lett.}\ }\textbf
  {\bibinfo {volume} {112}},\ \bibinfo {pages} {057202} (\bibinfo {year}
  {2014})}\BibitemShut {NoStop}%
\bibitem [{\citenamefont {Bandyopadhyay}\ \emph {et~al.}(2025)\citenamefont
  {Bandyopadhyay}, \citenamefont {Picozzi},\ and\ \citenamefont
  {Bhowal}}]{Bandyopadhyay2025}%
  \BibitemOpen
  \bibfield  {author} {\bibinfo {author} {\bibfnamefont {S.}~\bibnamefont
  {Bandyopadhyay}}, \bibinfo {author} {\bibfnamefont {S.}~\bibnamefont
  {Picozzi}},\ and\ \bibinfo {author} {\bibfnamefont {S.}~\bibnamefont
  {Bhowal}},\ }\href {https://arxiv.org/abs/2503.17001} {\bibinfo {title}
  {Designing non-relativistic spin splitting in oxide perovskites}} (\bibinfo
  {year} {2025}),\ \Eprint {https://arxiv.org/abs/2503.17001} {arXiv:2503.17001
  [cond-mat.mtrl-sci]} \BibitemShut {NoStop}%
\bibitem [{\citenamefont {Cricchio}(2010)}]{Cricchio2010}%
  \BibitemOpen
  \bibfield  {author} {\bibinfo {author} {\bibfnamefont {F.}~\bibnamefont
  {Cricchio}},\ }\emph {\bibinfo {title} {Multipoles in {Correlated} {Electron}
  {Materials}}},\ \href
  {http://urn.kb.se/resolve?urn=urn:nbn:se:uu:diva-132068} {Ph.D. thesis},\
  \bibinfo  {school} {Uppsala University} (\bibinfo {year} {2010})\BibitemShut
  {NoStop}%
\bibitem [{\citenamefont {Grånäs}(2012)}]{Granas2012}%
  \BibitemOpen
  \bibfield  {author} {\bibinfo {author} {\bibfnamefont {O.}~\bibnamefont
  {Grånäs}},\ }\emph {\bibinfo {title} {Theoretical {Studies} of {Magnetism}
  and {Electron} {Correlation} in {Solids}}},\ \href
  {http://urn.kb.se/resolve?urn=urn:nbn:se:uu:diva-172334} {Ph.D. thesis},\
  \bibinfo  {school} {Uppsala University} (\bibinfo {year} {2012})\BibitemShut
  {NoStop}%
\bibitem [{\citenamefont {Spaldin}\ \emph {et~al.}(2013)\citenamefont
  {Spaldin}, \citenamefont {Fechner}, \citenamefont {Bousquet}, \citenamefont
  {Balatsky},\ and\ \citenamefont {Nordstr\"om}}]{Spaldin2013}%
  \BibitemOpen
  \bibfield  {author} {\bibinfo {author} {\bibfnamefont {N.~A.}\ \bibnamefont
  {Spaldin}}, \bibinfo {author} {\bibfnamefont {M.}~\bibnamefont {Fechner}},
  \bibinfo {author} {\bibfnamefont {E.}~\bibnamefont {Bousquet}}, \bibinfo
  {author} {\bibfnamefont {A.}~\bibnamefont {Balatsky}},\ and\ \bibinfo
  {author} {\bibfnamefont {L.}~\bibnamefont {Nordstr\"om}},\ }\bibfield
  {title} {\bibinfo {title} {Monopole-based formalism for the diagonal
  magnetoelectric response},\ }\href
  {https://doi.org/10.1103/PhysRevB.88.094429} {\bibfield  {journal} {\bibinfo
  {journal} {Phys. Rev. B}\ }\textbf {\bibinfo {volume} {88}},\ \bibinfo
  {pages} {094429} (\bibinfo {year} {2013})}\BibitemShut {NoStop}%
\bibitem [{\citenamefont {Verbeek}\ \emph {et~al.}(2024)\citenamefont
  {Verbeek}, \citenamefont {Voderholzer}, \citenamefont {Sch\"aren},
  \citenamefont {Gachnang}, \citenamefont {Spaldin},\ and\ \citenamefont
  {Bhowal}}]{Verbeek2024}%
  \BibitemOpen
  \bibfield  {author} {\bibinfo {author} {\bibfnamefont {X.~H.}\ \bibnamefont
  {Verbeek}}, \bibinfo {author} {\bibfnamefont {D.}~\bibnamefont
  {Voderholzer}}, \bibinfo {author} {\bibfnamefont {S.}~\bibnamefont
  {Sch\"aren}}, \bibinfo {author} {\bibfnamefont {Y.}~\bibnamefont {Gachnang}},
  \bibinfo {author} {\bibfnamefont {N.~A.}\ \bibnamefont {Spaldin}},\ and\
  \bibinfo {author} {\bibfnamefont {S.}~\bibnamefont {Bhowal}},\ }\bibfield
  {title} {\bibinfo {title} {Nonrelativistic ferromagnetotriakontadipolar order
  and spin splitting in hematite},\ }\href
  {https://doi.org/10.1103/PhysRevResearch.6.043157} {\bibfield  {journal}
  {\bibinfo  {journal} {Phys. Rev. Res.}\ }\textbf {\bibinfo {volume} {6}},\
  \bibinfo {pages} {043157} (\bibinfo {year} {2024})}\BibitemShut {NoStop}%
\bibitem [{\citenamefont {Hoyer}\ \emph {et~al.}(2025)\citenamefont {Hoyer},
  \citenamefont {Stavropoulos}, \citenamefont {Razpopov}, \citenamefont
  {Valentí}, \citenamefont {Šmejkal},\ and\ \citenamefont
  {Mook}}]{Hoyer2025}%
  \BibitemOpen
  \bibfield  {author} {\bibinfo {author} {\bibfnamefont {R.}~\bibnamefont
  {Hoyer}}, \bibinfo {author} {\bibfnamefont {P.~P.}\ \bibnamefont
  {Stavropoulos}}, \bibinfo {author} {\bibfnamefont {A.}~\bibnamefont
  {Razpopov}}, \bibinfo {author} {\bibfnamefont {R.}~\bibnamefont {Valentí}},
  \bibinfo {author} {\bibfnamefont {L.}~\bibnamefont {Šmejkal}},\ and\
  \bibinfo {author} {\bibfnamefont {A.}~\bibnamefont {Mook}},\ }\href
  {https://arxiv.org/abs/2503.11623} {\bibinfo {title} {Altermagnetic splitting
  of magnons in hematite ($\alpha$-fe$_2$o$_3$)}} (\bibinfo {year} {2025}),\
  \Eprint {https://arxiv.org/abs/2503.11623} {arXiv:2503.11623
  [cond-mat.str-el]} \BibitemShut {NoStop}%
\bibitem [{\citenamefont {King-Smith}\ and\ \citenamefont
  {Vanderbilt}(1993)}]{Berry_phase1}%
  \BibitemOpen
  \bibfield  {author} {\bibinfo {author} {\bibfnamefont {R.~D.}\ \bibnamefont
  {King-Smith}}\ and\ \bibinfo {author} {\bibfnamefont {D.}~\bibnamefont
  {Vanderbilt}},\ }\bibfield  {title} {\bibinfo {title} {Theory of polarization
  of crystalline solids},\ }\href {https://doi.org/10.1103/PhysRevB.47.1651}
  {\bibfield  {journal} {\bibinfo  {journal} {Phys. Rev. B}\ }\textbf {\bibinfo
  {volume} {47}},\ \bibinfo {pages} {1651} (\bibinfo {year}
  {1993})}\BibitemShut {NoStop}%
\bibitem [{\citenamefont {Resta}(1993)}]{Berry_Phase2}%
  \BibitemOpen
  \bibfield  {author} {\bibinfo {author} {\bibfnamefont {R.}~\bibnamefont
  {Resta}},\ }\bibfield  {title} {\bibinfo {title} {Macroscopic electric
  polarization as a geometric quantum phase},\ }\href
  {https://doi.org/10.1209/0295-5075/22/2/010} {\bibfield  {journal} {\bibinfo
  {journal} {Europhysics Letters}\ }\textbf {\bibinfo {volume} {22}},\ \bibinfo
  {pages} {133} (\bibinfo {year} {1993})}\BibitemShut {NoStop}%
\bibitem [{\citenamefont {Resta}(1994)}]{Berry_Phase3}%
  \BibitemOpen
  \bibfield  {author} {\bibinfo {author} {\bibfnamefont {R.}~\bibnamefont
  {Resta}},\ }\bibfield  {title} {\bibinfo {title} {Macroscopic polarization in
  crystalline dielectrics: the geometric phase approach},\ }\href
  {https://doi.org/10.1103/RevModPhys.66.899} {\bibfield  {journal} {\bibinfo
  {journal} {Rev. Mod. Phys.}\ }\textbf {\bibinfo {volume} {66}},\ \bibinfo
  {pages} {899} (\bibinfo {year} {1994})}\BibitemShut {NoStop}%
\end{thebibliography}%

\vspace{1 cm}

\begin{center}
\large\bf{Supplemental Materials for \\ Rational Control of Magnonic and Electronic Band Splittings } 
\end{center}

\section{Computational Details}

The electronic structure of MnF$_2$, as presented in the manuscript, is computed using the plane-wave based projector augmented wave (PAW) \cite{Bloch1994, Kresse1999} method as implemented in the Vienna ab initio simulation package (VASP) \cite{Kresse1993, Kresse1996}. All the calculations are performed using the LDA+$U$ formalism with $U_{\rm eff}=5$ eV at the Mn site \cite{Yuan2020}. To achieve self-consistency, we use an energy cut-off of 550 eV and a $10\times10\times14$ k-point sampling of the Brillouin zone. The PAW potentials Mn-pv ([Mg]$3p^63d^54s^2$) and F ([He]$2s^22p^5$) are employed in the calculations. All the calculations are performed using the relaxed structure of MnF$_2$. The atomic relaxations are carried out until the Hellman-Feynman forces on each atom become less than 0.005 eV/\AA. The atomic-site multipoles at the Mn ions are calculated from the decomposition of the density matrix $\rho_{lm,l'm'}$, computed within the framework of density functional theory (DFT), into the tensor moments \cite{Cricchio2010, Granas2012, Spaldin2013}, of which the parity even tensor moments have contributions from the $l=l'$ terms.  

\begin{figure*}[ht]
\includegraphics[scale=0.65]{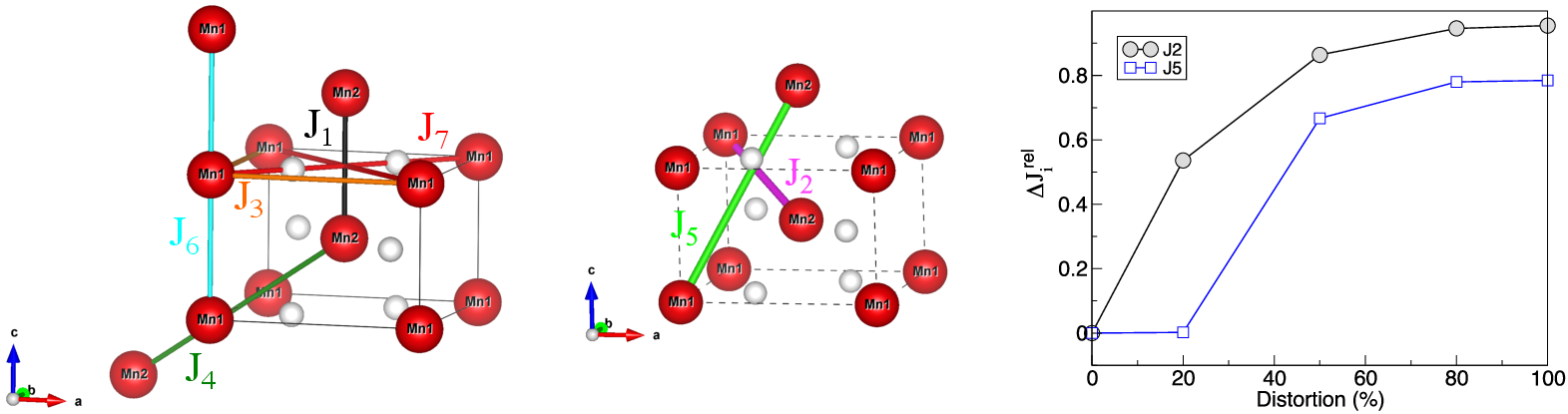}
\caption{ 
Magnetic exchange interaction paths $J_i$, with $i$=1 to 7, between the Mn atoms. The left and middle panels show the inter-sublattice and intra-sublattice Mn–Mn interaction paths, respectively. 
The right panel shows the variation of $\Delta J_i^{\rm rel}$ for $i=2$ and 5 as a function of amplitude of distortion.} 
\label{SM_all_J}
\end{figure*}

The exchange coupling $J_i$ between two Mn atoms at different distances is computed using the methodology described in Refs. \cite{Xiang2011, Sabani2020}. We have calculated $J_i$ up to the seventh nearest neighbor (NN). To calculate the exchange coupling between two particular Mn atoms, we compute the total energy within LDA+$U$ for four different collinear magnetic configurations while other Mn atoms are fixed to a ferromagnetic (FM) alignment.
While calculating $J$, we used an adequate size of supercells to isolate the chosen two Mn atoms (between which the exchange coupling is to be calculated) and avoid undesired contributions from other neighbors. Accordingly, [$1\times1\times4$], [$1\times1\times4$], [$2\times2\times2$], [$4\times2\times2$], [$3\times3\times3$], [$2\times2\times4$], [$2\times2\times5$], [$4\times4\times2$] supercells are used to calculate $J_i$s from first to seventh neighbours respectively. 
As the values of $J$ are very small, we paid special attention to reduce numerical error while calculating them. We set a tolerance of 10$^{-8}$ eV for the energy convergence during the electronic self-consistent calculations, which provided us the desired accuracy.
We also calculate the easy-axis anisotropy energy $D_c\sum_{i} (S_i^z)^2$ by computing the energy difference between AFM configurations with spin polarization along $\hat x$ and $\hat z$ in the presence of spin-orbit coupling. The computed value of $D_c$ is 1.2 $\mu$eV.

\begin{table}[h]
\caption{Frequencies of the optical phonon modes at the $\Gamma$ point}
\begin{tabular}{|c|c|c|c|}
\hline
$~$index$~$ & $~$frequency (cm$^{-1}$) & $~$index$~$ &$~$ frequency (cm$^{-1}$)\\
\hline
\hline
4 & 63 & 12 & 247\\
\hline
5 & 119 & \bf{13} & \bf{284}\\
\hline
6 & 157 &14 & 318\\
\hline
7 & 157 &\bf{15} & \bf{325}\\
\hline
8 & 220 &16 & 348 \\
\hline
9 & 234 &17 & 348 \\
\hline
10 & 234 & 18 & 453 \\
\hline
11 & 247 & & \\
\hline
\end{tabular}
\label{freq}
\end{table} 
The phonon band structure for the ground state $P4_2/mnm$ structure of MnF$_2$ is computed using the finite difference method as implemented within the PHONOPY software~\cite{Phonopy}. 
The $P4_2/mnm$ structure of MnF$_2$ has six atoms in the unit cell. For the phonon band structure calculation, we considered a $2\times2\times2$ supercell, keeping the ground state antiferromagnetic (AFM) spin alignment. No unstable phonon mode is found in our calculation, confirming the dynamic stability of the $P4_2/mnm$ structure. 
Frequencies of the 15 optical phonon modes at the $\Gamma$ point are given in Table \ref{freq}.

To obtain the high-energy $P4_2/nnm$ structure with zero spin and magnon splitting, we first construct the other structural domain with the same space group symmetry $P4_2/mnm$ but of opposite spin splitting
 \cite{BhowalSpaldin2024}.  The structure $P4_2/nnm$ is then obtained by linearly interpolating the two structures with opposite spin splitting. 
Distortions of the $P4_2/nnm$ structure can be described by the $\Gamma$ point phonon modes of the ground state $P4_2/mnm$ structure, which are identified using the following two steps. First, we calculate the atomic distortions of $P4_2/nnm$ relative to the $P4_2/mnm$ structure. This provides distortions in all three Cartesian directions. Next, these distortions are mapped to the phonon eigenvectors at the $\Gamma$ point as given below, 
\begin{equation}
    u_x\hat{x}+u_y\hat{y}+u_z\hat{z} = \sum_i c_i \vec{\Gamma_i}
    \label{map}
\end{equation}
Here, $u_j$ with $j=x, y, z$ are the atomic distortions, and $c_i$s are the amplitude of the  $\Gamma$ point phonon $i$ ($i=1,..,24$) mode with eigenvector $\vec{\Gamma}_i$.
We find that the reference structure, $P4_2/nnm$, with zero spin splitting contains distortion of the two stable $\Gamma$-phonon modes lying at  284 cm$^{-1}$(8.52 THz) and 325 cm$^{-1}$ (9.74 THz) respectively with $A_{2u}$ and $A_{1g}$ irreducible representations (IRs), indexed as optical phonon modes 13 and 15 in Table \ref{freq}. We normalize $c_{A_{2u}}$ and $c_{A_{1g}}$ to 1, which correspond to atomic distortions of  0.83\AA $~$ and 0.38\AA $~$  respectively as calculated using ISODISTORT\cite{Isodistort,Isodistort1}.

\section{Additional results of the exchange couplings}

\begin{table}[h!]
\caption{
Mn-Mn bond distances for the magnetic exchange interactions $J_i$. }
\begin{tabular}{|c|c|}
\hline
\hline
$~~~J_i~~~$ & Mn-Mn distance (\AA)  \\[0.5 ex]
\hline
\hline
$J_1$ & 3.35 \\
\hline
$J_2$ & 3.89 \\
\hline
$J_3$ & 4.97 \\
\hline
$J_4$ & 5.99 \\
\hline
$J_5$ &6.13 \\
\hline
$J_6$ &6.70  \\
\hline
$J_7$ &7.03\\
\hline
\hline
\end{tabular}
\label{SMtab2}
\end{table}

\begin{figure}[t]
\includegraphics[width=\columnwidth]{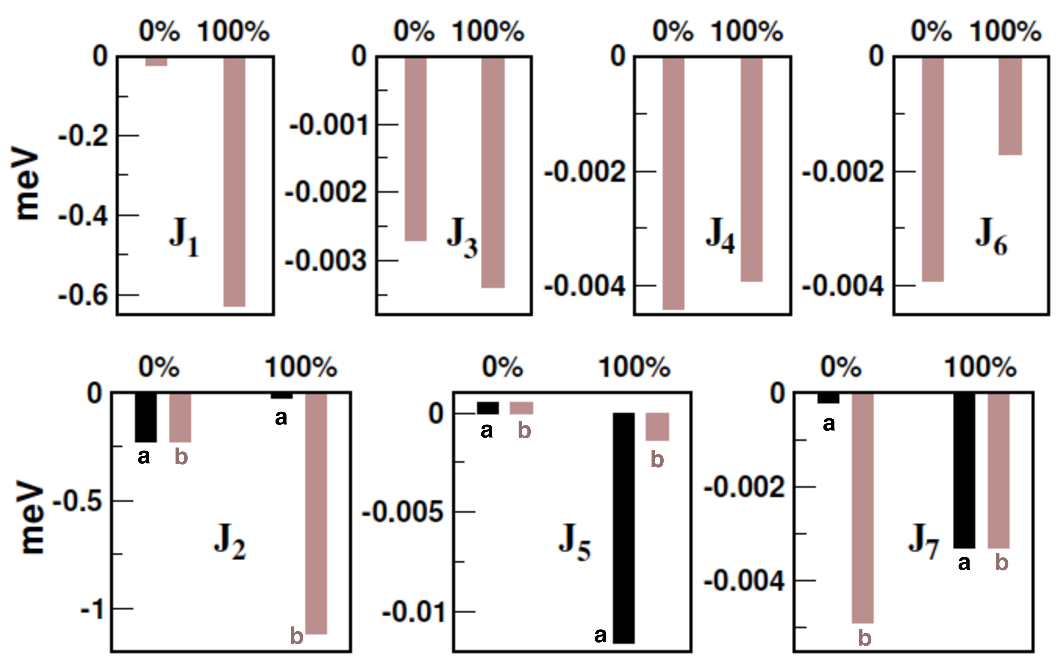}
\caption{Comparison of magnetic exchange interactions $J_i$ between the $P4_2/mnm$ (0\%) and $P4_2/nnm$ (100\%) structures. }
\label{SM_j}
\end{figure}

The exchange couplings $J_i$ up to the seventh NNs for the ground state $P4_2/mnm$ structure are indicated in Fig. \ref{SM_all_J} (a-b) and their computed values are shown in Fig. \ref{SM_j}. The corresponding distances are also listed in Table \ref{SMtab2}.
$\Delta J_7$ vanishes for $P4_2/nnm$ structure, explaining the suppression of the magnon band splitting. We note that the $J_2$ and $J_5$ exchange couplings, in contrast, become inequivalent for the $P4_2/nnm$ structure. We further compute $\Delta J_i^{\rm rel}= |\frac{J_i^a-J_i^b}{J_i^a+J_i^b}|$ with $i=2, 5$ for intermediate structures with different amplitudes of distortion, and the results are shown in Fig. \ref{SM_all_J}c. $\Delta J_i^{\rm rel}$ with $i=2, 5$, however, does not contribute to the splitting between magnon bands. \\

\begin{figure*}[ht]
\includegraphics[scale=0.45]{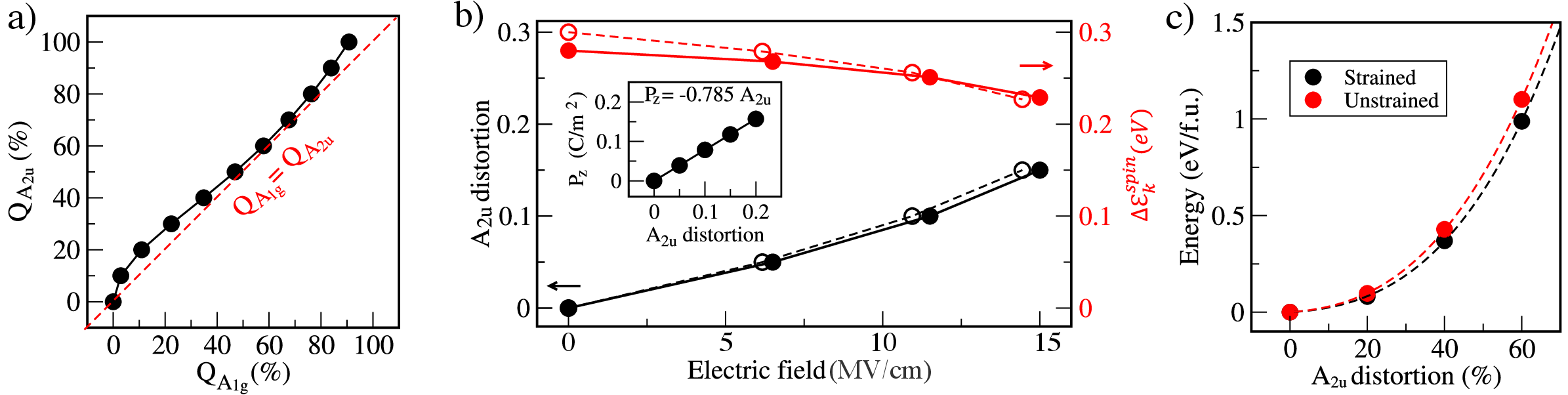}
\caption{ (a) Variation of $Q_{{\rm A}_{2u}}$ as a function of $Q_{{\rm A}_{1g}}$. Black line and dots represent the calculated optimal values of $Q_{{\rm A}_{2u}}$ for fixed values of $Q_{{\rm A}_{1g}}$. Red dashed line indicates  $Q_{{\rm A}_{1g}}=Q_{{\rm A}_{2u}}$. 
(b) Variation of the amplitude of static polar A$_{2u}$ distortion $Q_{{\rm A}_{2u}}$
induced by the electric field are shown with black solid and dashed lines for unstrained and strained structures, respectively. The
red solid and dashed lines show the corresponding spin-splitting energy of the second-highest valence bands for the unstrained
and strained structures. 
The inset shows the calculated electric polarization as a function of static A$_{2u}$ distortions. (c) Energy well of  $Q_{{\rm A}_{2u}}$, for unstrained and 2\% strained structures.}
\label{figsn1n2}
\end{figure*}

 \section{Details of the magnon spectrum }

In the main text, we analytically demonstrated that the direction-dependent $J_7$ exchange coupling is responsible for the splitting between the magnon modes in MnF$_2$, consistent with previous numerical results for the isostructural RuO$_2$ \cite{Libor2023}.
 Here we present more details in obtaining the analytical solution for the magnon dispersion, including the explicit forms of $A_k$, $B_k$, and $C_k$ from Eq. (1) in the main text. We use the Holstein-Primakoff transformation \cite{HolsteinPrimakoff1940} to map the spin operator, S, to bosonic creation (annihilation) operators $b_{i}^{\dagger}(b_{i})$ and $a_{i}^{\dagger}(a_{i})$ on the two sublattices. Here, the creation of the `$a$' (`$b$') boson at lattice site $i$ corresponds to reducing (increasing) the magnitude of the spin moment from $S$ to $S-1$  (-$S$ to -$S+1$) under the assumption $S \gg 1$, relevant to Mn$^{2+}: S=5/2$ in MnF$_2$. Using the Fourier transform of the operators, viz., $b_{k}^{\dagger} = \frac{1}{\sqrt{N}} \sum_{i} e^{ik.r_{i}} b_{i}^{\dagger}$, $b_{k} = \frac{1}{\sqrt{N}} \sum_{i} e^{-ik.r_{i}} b_{i}$, and similarly for the operator $a_{j}^{\dagger}(a_{j})$, we obtain Eq. (1) of the main text with the following forms $A_k$, $B_k$, and $C_k$ in terms of the various exchange couplings $J_i$ ($i = 1\text{–}7$):

\begin{widetext}

\begin{equation} 
\begin{aligned}
    A_{k} = &\, 2J_{1} \left[\cos(k_{z}c)-1\right] + 8J_{2} +2J_{3} \left[\cos(k_{y}a)+\cos(k_{x}a)-2\right] 
    + 2J_{4} \left[\cos(k_{z}c+k_{x}a)+\cos(k_{z}c-k_{x}a)-2\right]  \\
    &\,+24J_{5}+ 2J_{6} \left[\cos(k_{z}c)-1\right] + 2J_{7}^{a} \left[\cos(k_{y}+k_{x})a-1\right] + 2J_{7}^{b} \left[\cos(k_{y}-k_{x})a-1\right] 
\end{aligned}
\end{equation}

\begin{equation}
\begin{aligned}
    B_{k} = &\, 2J_{1} \left[\cos(k_{z}c)-1\right] + 8J_{2} +2J_{3} \left[\cos(k_{y}a)+\cos(k_{x}a)-2\right]  
   + 2J_{4} \left[\cos(k_{z}c+k_{x}a)+\cos(k_{z}c-k_{x}a)-2\right] \\
    &\,+24J_{5} + 2J_{6} \left[\cos(k_{z}c)-1\right] 
    + 2J_{7}^{a} \left[\cos(k_{y}-k_{x})a-1\right] + 2J_{7}^{b} \left[\cos(k_{y}+k_{x})a-1\right] 
\end{aligned}
\end{equation}
and

\begin{equation}
\begin{aligned}
    C_{k} &= C_{k}^{*} =  8 \cos\left(\frac{k_{x}a}{2}\right) \cos\left(\frac{k_{y}a}{2}\right) \cos\left(\frac{k_{z}c}{2}\right) \\
    &\quad \times \left[ J_2 + J_5 \left( 4\left( \cos^2\left(\frac{k_{x}a}{2}\right) + \cos^2\left(\frac{k_{y}a}{2}\right) + \cos^2\left(\frac{k_{z}c}{2}\right) \right) - 9 \right) \right]
\end{aligned}
\end{equation}
Here, $a$ and $c$ represent the tetragonal lattice constants of the MnF$_2$ unit cell. $J_i$ denotes the exchange coupling between the $i^{\rm th}$ NNs of Mn atoms.
\end{widetext}

\textcolor{black}{\section{Effective model}}

Symmetry adapted free energy expansion with $Q_{{\rm A}_{1g}}$ and $Q_{{\rm A}_{2u}}$ distortions is given by,
\begin{eqnarray}
    U &\approx & \alpha_1Q_{{\rm A}_{1g}}^2+  \alpha_2Q_{{\rm A}_{1g}}^4+ \beta_1Q_{{\rm A}_{2u}}^2+  \beta_2Q_{{\rm A}_{2u}}^4
     \nonumber\\
   & + & \lambda_1 Q_{{\rm A}_{2u}}^2Q_{{\rm A}_{1g}} + \lambda_2 Q_{{\rm A}_{2u}}^2Q_{{\rm A}_{1g}}^2+ O(Q^5)
   \label{freeenergy_SM}
\end{eqnarray}
We fit DFT energies to estimate the values of the coefficients in Eq.~(\ref{freeenergy_SM}). We used a DFT training set of 40 different  
structures in the fitting process. These structures are generated by condensing different combinations of $Q_{{\rm A}_{2u}}$ and $Q_{{\rm A}_{1g}}$. Optimal values of the Raman-active mode amplitude $Q_{{\rm A}_{1g}}$ for fixed values of the polar distortion $Q_{{\rm A}_{2u}}$ are estimated from Eq.~(\ref{freeenergy_SM}) by setting $\frac{dU}{dQ_{{\rm A}_{1g}}} = 0$. The obtained values of $Q_{{\rm A}_{1g}}$, as depicted in Fig.~\ref{figsn1n2}a as a function of $Q_{{\rm A}_{2u}}$, are close to the $Q_{{\rm A}_{2u}} = Q_{{\rm A}_{1g}}$ values.

Further, we estimate the required electric field to induce $Q_{{\rm A}_{2u}}$ polar distortions by including an extra term $\Omega$ ${\cal E}_z$$P_z$ in Eq. (\ref{freeenergy_SM}), as given below,
\begin{eqnarray}\label{electric_SM}
    U &\approx & \alpha_1Q_{{\rm A}_{1g}}^2+  \alpha_2Q_{{\rm A}_{1g}}^4+ \beta_1Q_{{\rm A}_{2u}}^2+  \beta_2Q_{{\rm A}_{2u}}^4
    + \lambda_1 Q_{{\rm A}_{2u}}^2Q_{{\rm A}_{1g}} 
    \nonumber\\
    &+& \lambda_2 Q_{{\rm A}_{2u}}^2Q_{{\rm A}_{1g}}^2+ O(Q^5)-\Omega ~{{\cal E}_z} P_z
\end{eqnarray}

Here, $\Omega$, ${\cal E}_z$, $P_z$ are respectively the volume of the unit cell, applied electric field, and the electric polarization of the system along $\hat z$. Here, $\Omega$ is 82.748 \AA$^3$. Further, we calculate the electric
polarization of the structures at different amplitudes of $Q_{{\rm A}_{2u}}$ polar distortions employing the Berry Phase method \cite{Berry_phase1,Berry_Phase2,Berry_Phase3}. The electric
polarization varies linearly with $Q_{{\rm A}_{2u}}$, viz., $P_z=0.785 Q_{{\rm A}_{2u}}$ C/m$^2$. Substituting $P_z$ in Eq. (\ref{electric_SM}) and solving the equation, we obtain the optimal value of ${\cal E}_z$ required to generate a given amplitude of $Q_{{\rm A}_{2u}}$ distortion, where we constraint  $Q_{{\rm A}_{2u}}=Q_{{\rm A}_{1g}}$. The results of our calculations are shown in Fig. \ref{figsn1n2}b, suggesting the possibility of achieving control over spin splitting with a reasonable external electric field \cite{Kyung_2023, Varignon_2016, Bellaiche_2014}. 

Further, we consider the effect of compressive biaxial strain along the $ab$ plane, which softens the $A_{2u}$ phonon mode, and consequently reduces the magnitude of $E_z$, required to tune the spin splitting. 
Our phonon calculation for the strained $P4_2/mnm$ structure with AFM configuration shows a reduction of 7.7\% in the frequency of the $A_{2u}$ phonon mode compared to the unstrained case. 
A softer $A_{2u}$ distortion leads to a flatter energy well, as shown in Fig. \ref{figsn1n2}c, resulting in a 4.5\% reduction in the harmonic coefficient ($\beta_1$) compared to the unstrained case. With the new coefficient $\beta_1$ and assuming other parameters unchanged, we estimate ${\cal E}_z$ from Eq. (\ref{electric_SM}). Our calculation shows a reduction in the value of ${\cal E}_z$ to induce a similar $Q_{A_{2u}}$ polar distortion amplitude, as shown in Fig. \ref{figsn1n2}b. 
Although this is a rough estimation, and in principle, the effect of strain on the other parameters in Eq. (\ref{electric_SM}) should also be considered, our calculations show that under bi-axial strain, the static combined $Q_{A_{2u}}$ and $Q_{A_{1g}}$ distortions lead to a modification of the spin splitting, which can now be achieved with a smaller applied electric field, as illustrated in Fig. \ref{figsn1n2}b.

\end{document}